\documentclass[a4paper, 11pt]{article}
\usepackage{geometry}
\usepackage{booktabs}
\usepackage{multirow}
\usepackage{footnote}
\usepackage[shortlabels]{enumitem}
\usepackage{float}
\usepackage{comment}
\usepackage{amsmath, amssymb}
\usepackage{amsfonts}
\usepackage{titlesec}
\usepackage{bm}
\usepackage{lscape} 
\usepackage{graphicx,comment}
\usepackage{url}
\usepackage{color}
\usepackage{xargs, xstring}
\usepackage{verbatim}
\usepackage{setspace} 
\usepackage[round]{natbib} 
\usepackage{algorithm}
\usepackage{bbm}
\usepackage[table, dvipsnames]{xcolor}
\usepackage{enumitem}
\usepackage{rotating}
\usepackage{tikz}
\usetikzlibrary{shapes.geometric, arrows}
\usetikzlibrary{automata,positioning}
\usepackage{multirow}
\def\bSig\boldsymbol{\Sigma}

\newcommand{\ind}{\perp \!\!\! \perp}
\usepackage{algpseudocode}

\usepackage{url}
\usepackage{cellspace}
\usepackage{subcaption}
\usepackage[colorlinks=true, linkcolor=black, urlcolor = black, citecolor=black]{hyperref} 
\usepackage[noabbrev,capitalize]{cleveref}
\usepackage[parfill]{parskip}    
\usepackage[pagewise]{lineno}

\usepackage{multicol}
\usepackage{array}
\usepackage{caption}

\usepackage{tabularx}
\usepackage{longtable}
\usepackage{authblk}

\usepackage[export]{adjustbox} 

\usepackage{comment} 

\usepackage{microtype}              
\usepackage{ragged2e}  
\usepackage{tabularray}
\UseTblrLibrary{booktabs}           
\NewTableCommand\category[1][0pt]{
        \SetRow{abovesep+=#1}
        \SetCell[c=4]{l, font=\footnotesize\itshape\bfseries}
    \SetTblrStyle{contfoot-text}{font=\footnotesize\itshape}                            }

\providecommand{\keywords}[1]
{
  \small	
  \textbf{\textit{Keywords---}} #1
}
\newcommand{\be}{\begin{eqnarray}}
\newcommand{\ee}{\end{eqnarray}}
\newcommand{\bee}{\begin{eqnarray*}}
\newcommand{\eee}{\end{eqnarray*}}
\newcommand{\bi}{\begin{enumerate}}
\newcommand{\ei}{\end{enumerate}}

\crefformat{equation}{(#2#1#3)}

\title{Representative dietary behavior patterns and associations with cardiometabolic outcomes in Puerto Rico using a Bayesian latent class analysis for non-probability samples}
\author{Stephanie M. Wu $^{1,*}$, 
Abrania Marrero $^{2}$,
Matthew R. Williams $^{3}$, 
Terrance D. Savitsky $^{4}$, 
Josiemer Mattei $^{2}$,
José Rodríguez-Orengo $^{5}$
Briana J.K. Stephenson $^{1}$
\\

\footnotesize $^{1}$Department of Biostatistics, Harvard T.H. Chan School of Public Health, Boston, Massachusetts, U.S.A \\
\footnotesize $^{2}$Department of Nutrition, Harvard T.H. Chan School of Public Health, Boston, Massachusetts, U.S.A \\
\footnotesize $^{3}$RTI International, Research Triangle Park, North Carolina, U.S.A\\
\footnotesize $^{4}$Office of Survey Methods Research, U.S. Bureau of Labor Statistics, Washington, DC, U.S.A\\
\footnotesize $^{5}$Department of Biochemistry, Medical Sciences Campus, University of Puerto Rico, San Juan, PR\\
\footnotesize *\textit{email:} swu@g.harvard.edu 
}
\date{} 

\begin{document}\emergencystretch 3em

\maketitle

\singlespacing

\begin{abstract}
There is limited understanding of how dietary behaviors cluster together and influence cardiometabolic health at a population level in Puerto Rico. Data availability is scarce, particularly outside of urban areas, and is often limited to non-probability sample (NPS) data where sample inclusion mechanisms are unknown. In order to generalize results to the broader Puerto Rican population, adjustments are necessary to account for selection bias but are difficult to implement for NPS data. Although Bayesian latent class models enable summaries of dietary behavior variables through underlying patterns, they have not yet been adapted to the NPS setting. We propose a novel Weighted Overfitted Latent Class Analysis for Non-probability samples (WOLCAN). WOLCAN utilizes a quasi-randomization framework to (1) model pseudo-weights for an NPS using Bayesian additive regression trees (BART) and a reference probability sample, and (2) integrate the pseudo-weights within a weighted pseudo-likelihood approach for Bayesian latent class analysis, while propagating pseudo-weight uncertainty into parameter estimation. A stacked sample approach is used to allow shared individuals between the NPS and the reference sample. We evaluate model performance through simulations and apply WOLCAN to data from the Puerto Rico Observational Study of Psychosocial, Environmental, and Chronic Disease Trends (PROSPECT). We identify dietary behavior patterns for adults in Puerto Rico aged 30 to 75 and examine their associations with type 2 diabetes, hypertension, and hypercholesterolemia. Our findings suggest that an out-of-home eating pattern is associated with a higher likelihood of these cardiometabolic outcomes compared to a nutrition-sensitive pattern. WOLCAN effectively reveals generalizable dietary behavior patterns and demonstrates relevant applications in studying diet-disease relationships.
\end{abstract}

\keywords{Bayesian model-based clustering, Quasi-randomization, Survey sampling, BART, Diet, Puerto Rico}


\section{Introduction}\label{sec:intro}
Substantial evidence has demonstrated the benefits of a healthy diet in the primary and secondary prevention of cardiometabolic conditions, such as hypertension, type 2 diabetes, hypercholesterolemia, and cardiovascular disease \citep{cespedes2015dietary, schwingshackl2015diet}. Valuable insights into the factors influencing dietary intake and subsequent health outcomes can be gained by analyzing dietary behaviors, including food purchasing behavior and dieting practices (e.g., controlling sugars, salt, and portion sizes in the diet). These behaviors are also embedded within unique psychosocial, environmental, cultural, and socioeconomic contexts \citep{kris2020barriers, marijn2018dietary}. While previous studies have explored the impact of individual or selective dietary behaviors on cardiometabolic outcomes \citep{hollar2010healthier, keyserling2016community, kelli2019living}, a comprehensive analysis examining dietary behavior patterns, incorporating multiple variables simultaneously, remains largely underrepresented in the literature. 

The limited availability of nutritional data often hinders efforts to study dietary behavior factors and their role in cardiometabolic health. This challenge is particularly evident in Puerto Rico (PR), where cardiometabolic diseases are highly prevalent, with type 2 diabetes affecting 17\% of the population, hypertension 42\%, and hypercholesterolemia 41\% \citep{centers2021brfss}. Despite these concerning statistics, population-level studies on the health and behaviors of Puerto Rican residents remain limited. Existing research is often confined to urban centers only \citep{perez2013prevalence, lopez2023self, suglia2024prevalence}, largely due to the logistical difficulties in obtaining more representative geographically diverse samples. 

The Puerto Rico Observational Study of Psychosocial, Environmental, and Chronic Disease Trends (PROSPECT) aims to bridge this gap by providing rich data on dietary behaviors and cardiometabolic outcomes across PR to better examine cardiovascular-related disparities that persist across class and geographies \citep{mattei2021design}. PROSPECT was initially designed with a multistage cluster probability sampling design. However, enrollment was severely hindered by issues of inaccurate household information in the aftermath of major population shifts resulting from the passage of Hurricane Maria, an economic recession, and relatively high levels of non-participation, potentially reflective of community mistrust in medical institutions \citep{santos2020puerto, davis2012sociodemographic}. To effectively adapt to these challenges, PROSPECT incorporated a combination of various community-based sampling techniques, partnering with local clinics and community groups, to bolster enrollment and ensure data were obtained from important demographic subpopulations. This allowed the sample to more closely reflect the age, sex, and regional composition of Puerto Rico. 

While convenient, this revised community-based sampling design results in a non-probability sample (NPS), where the sampling mechanism is not governed by known selection probabilities. As a result, unknown selection biases arising during the data collection process generate a sample composition that is not representative of the population composition. This complicates the ability to make inferences from this dataset, as it is not generalizable to the broader population of adults living in PR. Adjustments for selectivity and inclusion into PROSPECT must be made to avoid biased and misleading results when extending any analyses to the population level \citep{baker2013summary}. 

Population-level inference is typically reliant on probability sampling, where sampled individuals' selection probabilities are known \citep{rao2017sample}. However, probability samples are increasingly encumbered by high nonresponse rates, cost, and time-intensive implementation \citep{meyer2015household}. This has shifted attention towards non-probability samples, which employ strategies such as respondent-driven sampling and community-based sampling and do not require individuals to all be randomly selected into the sample \citep{heckathorn1997respondent, krueger2020comparing}. Non-probability samples are best suited for reaching rare, hard-to-reach, and stigmatized populations, and also offer time and cost advantages. Statistically, non-probability samples require particular care in building models that adjust for potential sources of bias that jeopardize generalization to the population. 

\subsection{Methods for Non-Probability Samples}
Existing methods for estimation and inference from non-probability samples have focused on univariate summary statistics and include, broadly speaking, quasi-randomization, prediction modeling, and doubly-robust approaches \citep{valliant2020comparing, elliott2017inference, wu2022statistical}.  In these approaches, selectivity into the NPS is modeled by borrowing information from a reference sample that is a probability sample (PS) with known inclusion probabilities. In quasi-randomization methods, also referred to as propensity score models, an unknown underlying random mechanism is assumed to govern selection into the NPS and is estimated through the creation of pseudo-weights for the NPS. To estimate these pseudo-weights, the known PS inclusion probabilities are combined with information differentiating the NPS units from the PS units. This requires a set of auxiliary covariates to be present in both samples in order to fully explain the sampling mechanism. A number of quasi-randomization approaches have been recently developed \citep{valliant2018nonprobability, wang2021adjusted, rafei2020big, savitsky2022methods}, with a review of such approaches provided in \citet{beresovsky2025review}. Alternative strategies for non-probability samples include prediction modeling, where the NPS is used to train a model for the outcome based on the auxiliary covariates and then outcomes for the PS are predicted using this outcome model \citep{kim2021combining, wang2015forecasting}, as well as doubly robust methods that combine quasi-randomization and prediction modeling \citep{chen2020doubly, rafei2022robust, liu2023inference}. Prediction modeling and doubly-robust approaches work well for univariate outcomes, but are difficult to implement for more complex models with multivariate outcomes, as is the case for our application to dietary behavior patterns. They are also better suited to settings with a large NPS, such as expansive web-based surveys, as opposed to the smaller sample size in PROSPECT necessitated by the comprehensiveness of the data collected. Consequently, we focus on quasi-randomization approaches for our setting. 

Current quasi-randomization approaches have not yet been extended to the multivariate and correlated data setting that we see in analyzing dietary behavior patterns. Latent class analysis (LCA; \citet{lazarsfeld1968latent}) encompasses a class of models that are able to summarize a large number of categorical and correlated behavior variables into underlying patterns and archetypes. However, the increased complexity of the data in LCA adds additional complications when adjusting for selection bias. It also makes outcome modeling more challenging, precludes the use of standard formulas for variance estimation, and greatly complicates how uncertainty in the estimation of the NPS pseudo-weights is incorporated. 

Adopting a Bayesian framework provides advantages for addressing these challenges, as parameter uncertainty can be captured through posterior samples rather than closed-form variance expressions. \citet{rafei2020big} and \citet{liu2023inference} explored the use of Bayesian additive regression trees (BART; \citet{chipman2010bart}), a flexible machine learning method that has high predictive performance and allows for complex interactions, to improve pseudo-weight estimation in non-probability samples and enable pseudo-weight uncertainty to be propagated. Furthermore, recent survey-weighted Bayesian LCA techniques enable elicitation of latent patterns, such as dietary behavior archetypes, that are generalizable beyond the sample to the broader population; however, they rely on known survey weights and survey design information to be available \citep{wu2024derivation, stephenson2024identifying}. In non-probability samples, such information on the selection mechanism is unknown and must be estimated through additional statistical modeling. We draw from these two areas to develop a Bayesian machine learning model-based clustering method that enables generalizable LCA for non-probability samples.

In this work, our goal is to use the data in PROSPECT, a community-based NPS, to make generalizable inferences to the broader Puerto Rican adult population on dietary behaviors and their impact on health outcomes. This is of public health importance because the health of adults living in Puerto Rico is understudied at the population level despite the high prevalence of cardiometabolic diseases \citep{mattei2021design}. Our main contributions are to: 1) extend Bayesian weighted LCA models to the NPS setting by using BART and a quasi-randomization framework to account for selection biases; 2) ensure accurate point and variance estimation through extensive simulation studies; and 3) derive representative dietary behavior patterns for the population of adults living in PR and examine how these patterns relate to type 2 diabetes, hypertension, and hypercholesterolemia.

The remainder of this article is organized as follows: Section \ref{sec:model_WOLCAN} describes the proposed model. Section \ref{sec:sims_WOLCAN} discusses an extensive simulation study to assess model performance and estimation under different settings. Section \ref{sec:application} implements our proposed model to the PROSPECT study to derive dietary behavior patterns of adults living in Puerto Rico and examine their relationships to cardiometabolic risk factors (type 2 diabetes, hypertension, hypercholesterolemia). Section \ref{sec:discussion} concludes with a discussion of strengths, limitations, and extensions.

\section{Model Formulation}\label{sec:model_WOLCAN}
The proposed model, referred from this point forward as Weighted Overfitted Latent Class Analysis for Non-probability samples (WOLCAN) is constructed in two steps. First, pseudo-sampling weights are generated for the non-probability sample using a quasi-randomization (QR) stacked sample approach with estimation carried out via Bayesian additive regression trees (BART). Second, the estimated pseudo-weights are incorporated into a Bayesian LCA framework to identify latent patterns across a large set of categorical and correlated variables. Detailed information on each of these steps is provided below. 

\subsection{Estimation of Pseudo-Sampling Weights Using BART}\label{subsec:pseudo_weights}
Let $U$ denote a population of size $N$. Consider an NPS $S_B$ with sample size $n_B$, and a reference PS $S_R$ with sample size $n_R$, both drawn from $U$. In the NPS, the selection mechanism into the sample is complex, unknown, and must be estimated. We utilize a QR framework where we assume there is some unknown, underlying sampling design giving rise to the NPS, which can be recovered through statistical modeling that borrows information from the reference PS. Indicator variables $\delta_i^B = I(i \in S_B)$ and $\delta_i^R = I(i \in S_R)$ denote individual selection into the NPS and the PS, respectively, equal to 1 if individual $i$ in the population is selected, and 0 otherwise. To obtain the pseudo-sampling weights, we need to estimate the \textit{unknown} individual inclusion probabilities for the NPS, denoted by $\pi_i^B = P(\delta_i^B = 1), \, i \in S_B$. This is achieved by borrowing information from the \textit{known} individual inclusion probabilities for the PS, denoted by $\pi_i^R = P(\delta_i^R = 1), \, i \in S_R$. Selection into the samples is determined by observed individual-level auxiliary variables, $\mathbf{a}_i$, available in both the NPS and the PS. The individual-level auxiliary variables are not required to be known for the entire population, in contrast to other methods \citep{liu2023inference, elliott2017inference}. 

The pseudo-weight estimation step requires the following assumptions:

\begin{enumerate}
    \item Ignorability: The selection mechanism for NPS $S_B$ is ignorable given the set of observed auxiliary variables $\mathbf{a}_i$. Letting $\boldsymbol{x}_i$ denote the multivariate observed data, we have: $P(\delta_i^B = 1|\mathbf{x}_i, \mathbf{a}_i) = P(\delta_i^B = 1|\mathbf{a}_i)$ for all $i \in U$.
    \item Positivity: All individuals have a positive probability of inclusion in $S_B$: $P(\delta_i^B = 1|\mathbf{a}_i) > 0$ for all $i \in U$.
    \item Pairwise independence: Individuals of $S_B$ are selected independently given the auxiliary variables: $\delta_i^B \ind \delta_j^B | \mathbf{a}_i $ for $i \neq j$ .
    \item Sampling independence: Sampling for $S_B$ and $S_R$ are independent given the auxiliary variables: $\delta_i^B \ind \delta_i^R | \mathbf{a}_i$ for all $i \in U$.
\end{enumerate}

\begin{figure}
    \centering
    \includegraphics[width=0.9\linewidth]{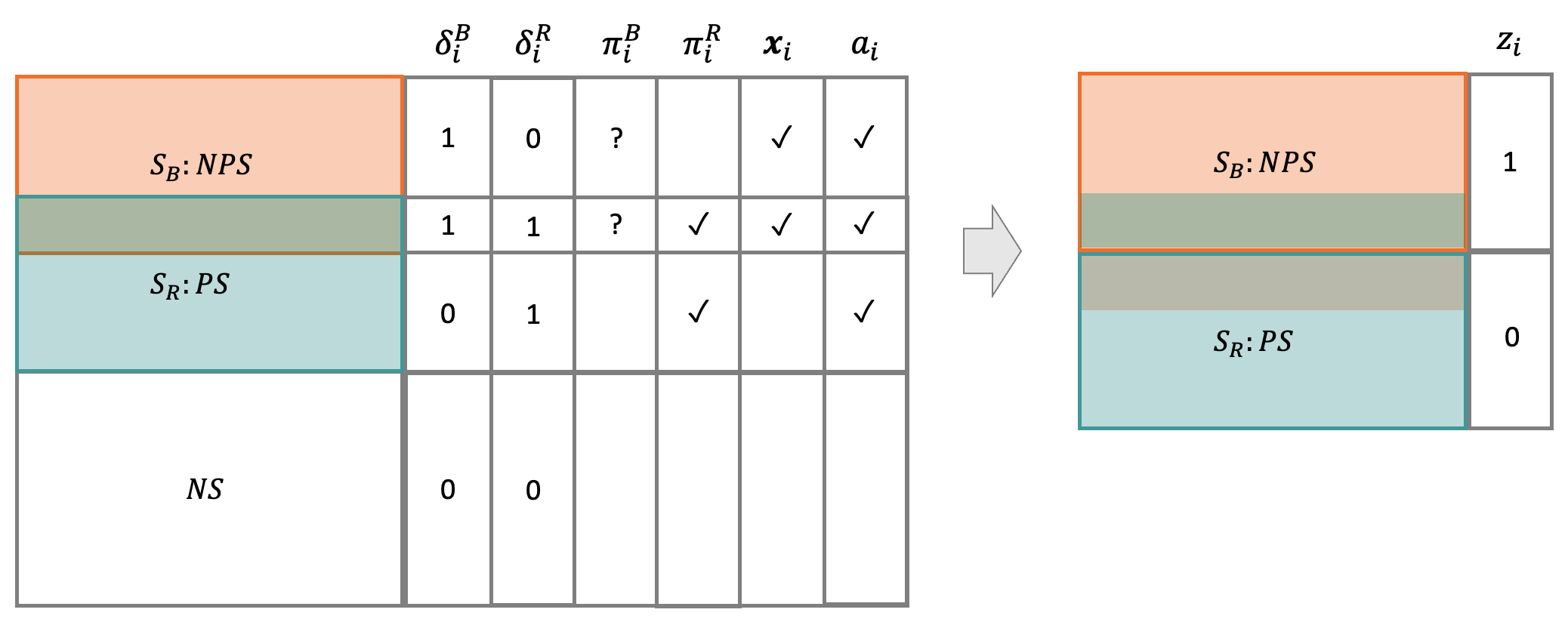}
    \caption{\textmd{Example data structure in the population (left) and the stacked sample (right) when there are overlapping individuals that belong to both the NPS and the PS. NS = non-sampled individuals. Check mark indicates the variable is available. Blank space or question mark indicates the variable is not available.}}
    \label{fig:overlap_diagram}
\end{figure}

We adopt the ``stacked sample'' approach from \citet{beresovsky2025review} and \citet{savitsky2022methods}, which accommodates overlap between the NPS and the reference PS. This approach broadens the method's applicability, allowing it to handle settings where individuals may or may not be shared between the NPS and PS, even when the identities of shared individuals are unknown. First, we stack the NPS and the PS together to form a combined sample $S$ with sample size $n = n_B + n_R$. Individuals belonging to both samples, also known as overlap individuals, are included twice. For each individual $i \in S$ in the combined sample, let $z_i$ denote the binary indicator equal to 1 if individual $i$ is in the NPS, and 0 otherwise, and let $\pi_i^Z = P(z_i = 1)$, $i \in S$, be the propensity of being in the NPS given inclusion in the combined sample, modeled using the auxiliary variables $
\mathbf{a}_i$. The stacked sample approach allows us to estimate the pseudo-inclusion probabilities for the NPS, $\pi_i^B$, by using the estimated propensity of being in the NPS rather than the PS, $\pi_i^Z$. It is advantageous over alternative approaches because it allows for the proper inclusion of overlap individuals without needing to identify who these individuals are, in contrast to approaches that require no overlap between samples \citep{rafei2020big, elliott2017inference}, and it is more efficient than approximate, pseudo-likelihood methods \citep{chen2020doubly, wang2021adjusted, valliant2011estimating}. Figure \ref{fig:overlap_diagram}  provides a simple diagram illustrating an example of the data structure in the population and stacked sample when there are overlapping individuals, assuming both sampling frames cover the population for simplicity. 

We construct a propensity score model describing $\pi_i^Z$, the probability that an individual belongs to the NPS given that they are included in the combined sample, using Bayesian additive regression trees (BART; \citet{chipman2010bart}) and following a similar construction as in \citet{rafei2020big}. BART is a flexible, non-parametric statistical method that models the relationship between covariates and an outcome by summing together many individual regression trees, which each contribute a small part to the overall prediction.  BART has strong predictive performance compared to standard regression methods and can handle complex and non-linear associations. It also quantifies uncertainty in predictions through Bayesian posterior predictive draws. In our setting, this enables uncertainty from estimating the pseudo-weights to be directly propagated to estimation of dietary behavior patterns because we are using a Bayesian model.

\begin{figure}
    \centering
    \includegraphics[width=0.3\linewidth]{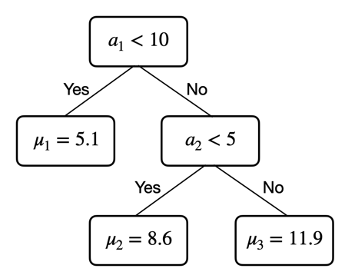}
    \caption{\textmd{Toy example diagram displaying a binary tree component of BART}}
    \label{fig:bart}
\end{figure}

Suppose we have $Q$ individual-level auxiliary variables, $\mathbf{a}_i=(a_{i1}, \ldots, a_{iQ})$, available for all individuals in the NPS and PS. We model selection into the samples dependent on these auxiliary variables and assume ignorable missingness conditional on them. The probability that an individual belongs to the NPS given that they are included in the combined sample, $\pi_i^Z$, can be estimated by fitting a probit BART model for binary outcomes:
\begin{equation}\label{eq:probit_BART}
    \pi_i^Z := E(z_i \mid \boldsymbol{a}_i) = \Phi\left\{\sum_{t=1}^{T} f(\boldsymbol{a}_i, T_t, \boldsymbol{\mu}_t)\right\}, \quad i \in S.
\end{equation}
$\Phi(\cdot)$ is the standard normal cumulative distribution function, $T_t$ is the $t$-th binary tree structure that partitions the input space into non-overlapping regions $\{R_{t1},\ldots, R_{tL_t}\}$, and $\boldsymbol{\mu}_t = \{ \mu_{t1}, \ldots, \mu_{tL_t} \}$ is the vector of respective predicted mean values at the $L_t$ terminal nodes for tree $T_t$. $f(\boldsymbol{a}_i, T_t, \boldsymbol{\mu}_t ) = \sum_{l=1}^{L_t}\mu_{tl}I(\boldsymbol{a}_i \in R_{tl})$ is the binary regression tree function that maps input $\boldsymbol{a}_i$ to a predicted mean value. Each tree is formed by a sequence of binary decision rules that split observations into more homogeneous subgroups and then assign an estimated value $\mu_i$ based on the associated terminal node. A toy example of $f(\boldsymbol{a}_i, T_t, \boldsymbol{\mu}_t )$ for a tree with three terminal nodes is pictured in Figure \ref{fig:bart}. The binary tree components are then combined in BART using a sum-of-trees model. As a Bayesian procedure, the model parameters have a regularization prior placed on them and are updated using a combination of the backfitting and MCMC Gibbs sampler methods. Further details of this sampler can be found in \citet{chipman2010bart}.

Though the PS inclusion probabilities $\pi_i^R$ are known for those in the PS, they are often not known for those in the NPS and must be estimated. We model $\pi_i^R$ for $i \in S_B$ with a logit transformation and use BART for continuous outcomes. Those in the PS belong to the training set and those in the NPS belong to the test set. 
\begin{equation}\label{eq:BART}
    \text{logit}(\pi_i^R) := \text{logit}\left\{E(\delta_i^R \mid \boldsymbol{a}_i)\right\} = \sum_{t=1}^{T} f(\boldsymbol{a}_i, T_t, \boldsymbol{\mu}_t), \quad i \in S_B,
\end{equation}
where $\boldsymbol{a}_i$, $T_t$, $\boldsymbol{\mu}_t $, and $f(\cdot)$ are defined as in Equation (\ref{eq:probit_BART}). 

By applying the BART models in Equations (\ref{eq:probit_BART}) and (\ref{eq:BART}), we can obtain respective estimates of $\hat{\pi}_i^Z$, the propensity of NPS assignment given inclusion in the combined sample, and of $\hat{\pi}_i^R$, the PS inclusion probabilities, for all individuals in the NPS (i.e., $i \in S_B$). Both BART models provide output in the form of $M$ posterior predictive samples, where $M$ is some large number such as 1000. For each iteration $m$ of the BART posterior predictive samples, we utilize the estimated quantities, $\hat{\pi}_i^{Z(m)}$ and $\hat{\pi}_i^{R(m)}$, and apply the CRISP formula from \citet{beresovsky2025review} to obtain the corresponding desired pseudo-inclusion probabilities for the NPS:
\begin{equation}
    \hat{\pi}_i^{B(m)} = \frac{\hat{\pi}_i^{Z(m)} \pi_i^R p_i^R}{p_i^B \big(1 - \hat{\pi}_i^{Z(m)}\big)}, \quad i \in S_B.
\end{equation}
$p_i^R$ and $p_i^B$ denote the known coverage proportion of the sampling frame for the PS and NPS, respectively. If all individuals in the population of interest have the potential to be sampled for both samples, then $p_i^R$ and $p_i^B$ are equal to 1; otherwise, the values are less than 1. The final pseudo-sampling weights for those in the NPS are obtained by inverting the pseudo-inclusion probabilities for those in the NPS:
\begin{equation}
    \hat{w}_i^{(m)} = \frac{1}{\hat{\pi}_i^{B(m)}}, \quad \text{with mean} \quad \hat{w}_i = \frac{1}{M} \sum_{m=1}^{M} \hat{w}_i^{(m)}, \quad i \in S_B.
\end{equation}

Noisiness in the estimation of the pseudo-weights can be reduced through approaches such as weight-trimming. We consider one ad-hoc method of trimming extreme weights where a cut-off value is specified based on the median plus a multiple $c$ of the interquartile range (IQR) \citep{chowdhury2007weight, potter2015methods}. Therefore, all weights are bounded by $ Q_2 + c\times  (Q_3 - Q_1)$, where $Q_2, Q_3,$ and $Q_1$ are the median, 3rd and 1st quartiles of the pseudo-weights, respectively, and $c$ is an arbitrary constant. Weight trimming procedures allow for a possible increase in bias in order to reduce noisy estimation due to highly extreme weights. Additional approaches can be used to calibrate the trimmed weights to known population benchmarks \citep{folsom2000generalized, williams2024optimization}.

\subsection{Bayesian Weighted Overfitted Latent Class Analysis for Non-Probability Samples}\label{subsec:WOLCAN}

After the pseudo-sampling weights are estimated, they can now be incorporated into the main model to derive generalizable dietary behavior patterns. We integrate the pseudo-sampling weights by using a weighted pseudo-likelihood approach \citep{savitsky2016bayesian}, which allows us to re-weight the individuals in the NPS to better reflect the target population. We introduce a weighted overfitted latent class analysis for non-probability samples (WOLCAN). WOLCAN extends the weighted overfitted latent class analysis (WOLCA; \citet{stephenson2024identifying, wu2024derivation}) for survey data to the NPS setting by incorporating the estimated pseudo-sampling weights described above.

Let $i \in \{1, \ldots, n_B\}$ index individuals in the NPS, and let $j \in \{1, \ldots, J\}$ index the exposure items of interest (e.g., dietary behaviors), which are only available in the NPS. Each exposure item $x_{ij}$ is categorical, following a Multinomial distribution with levels $r \in \{1,\ldots, R_j\}$ and item level probabilities $\theta_{jc_ir}$, where  $c_i \in \{1, \ldots, K\}$ is the latent class membership for individual $i$ and $K$ denotes the number of latent classes (i.e., patterns). The parameter $\pi_k$ denotes the probability of membership in class $k$, with $\sum_{k=1}^{K} \pi_k = 1$, $k \in \{1,\ldots, K\}$. The parameter $\theta_{jkr}$ characterizes the latent class pattern profiles and is the probability of level $r$ for variable $j$ and class $k$, with $\sum_{r=1}^{R_j} \theta_{jkr} = 1$, $j\in \{1,\ldots, J\}$, $k \in \{1,\ldots, K\}$. The individual complete data density follows the standard LCA form and is given by:
\begin{equation}
    p(\mathbf{x}_i, c_i \mid \pi, \theta) = \prod_{k=1}^{K} \left\{ \pi_k \prod_{j=1}^{J} \prod_{r=1}^{R_j} \theta_{jkr}^{I(x_{ij} = r)} \right\}^{I(c_i = k)}.
\end{equation}

LCA requires the following additional assumptions:
\begin{enumerate}
    \item  Local independence: The categorical variables are assumed to be mutually independent within each latent class: $x_{ij} \ind x_{il} | c_i$ for $j \neq l$.
    \item  Partitioning: Each individual is assumed to belong to one and only one latent class, and these latent classes are mutually exclusive and exhaustive: $\sum_{k=1}^K P(c_i=k|\mathbf{x}_i) = 1$.
\end{enumerate}

The number of patterns, $K$, is determined using a data-driven Bayesian overfitted formulation \citep{van2015overfitting}, where $K$ is set to be a conservatively high number to allow empty patterns to drop out via a sparsity-inducing Dirichlet prior: $(\pi_1, \ldots, \pi_K) \sim \text{Dir}(\alpha_1, \ldots, \alpha_K)$, where hyperparameters $\alpha_k$ moderate the rate of growth for nonempty classes and are recommended to be set to $1/K$. This is referred to as the \textit{adaptive sampler}, where the model is run specifically to obtain the estimated number of patterns, $\hat{K}$. 

Let $\hat{w}_i$ denote the mean pseudo-sampling weight for individual $i$, $i \in \{1, \ldots, n_B\}$, from Section \ref{subsec:pseudo_weights}. The pseudo-weights are incorporated into parameter estimation by up-weighting the likelihood contribution of each individual in the NPS proportional to the estimated number of individuals they represent in the target population. This forms a \textit{weighted pseudo-likelihood} that is used instead of the likelihood in the Bayesian posterior update. Let $\mathbf{\pi} = (\pi_1,\ldots, \pi_K)^T$ and $\mathbf{\theta}$ denote a $J\times K \times R$ array with cells $\theta_{jkr}$, where $R=\max_jR_j$. The posterior distribution of the parameters is updated proportional to the prior times the weighted pseudo-likelihood:
\begin{equation}\label{eq:pseudo_likelihood}
    p(\mathbf{\pi}, \mathbf{\theta}) \propto p(\mathbf{\pi}, \mathbf{\theta}) \prod_{i=1}^{n} \left\{ p(\mathbf{x}_i, c_i \mid \mathbf{\pi}, \mathbf{\theta})^{\frac{\hat{w}_i}{\kappa}} \right\},
\end{equation}
where $\kappa = \sum_{i=1}^{n_B} \hat{w}_i$ is the normalization constant used so that the weights sum to the sample size to account for sampling uncertainty. The parameters of interest are: the posterior class assignment probabilities, $\pi_k$, $k=1,\ldots, K$, characterizing the prevalence of each pattern in the population; and the posterior item level probabilities, $\theta_{jkr}$, $j=,1\ldots, J, k=1,\ldots, K, r = 1, \ldots, R$, characterizing the pattern-specific profiles. Estimation of parameters proceeds via an MCMC Gibbs sampler, with more details provided in \citet{wu2024derivation}. 

Since the pseudo-weights were estimated rather than known, uncertainty from this estimation must be accounted for during parameter estimation. To do this, we re-run model estimation using various draws from the distribution of pseudo-weights. We begin by drawing $D$ (e.g., 20) sets of pseudo-weights from the posterior predictive distribution obtained from the BART procedures, applying weight trimming as necessary. For each individual $i \in \{1,\ldots, n_B\}$, we have $(\hat{w}_i^{(1)}, \hat{w}_i^{(2)}, \ldots, \hat{w}_i^{(D)})$. We use evenly spaced quantiles of the pseudo-weight distribution to improve efficiency of this drawing process. For each set of pseudo-weights, $(\hat{w}_1^{(d)}, \ldots, \hat{w}_{n_B}^{(d)})$, $d=1, \ldots, D$, we normalize to sum to the sample size using normalization constant $\kappa^{(d)} = \sum_{i=1}^{n_B} \hat{w}_i^{(d)}$. The posterior distribution of the parameters is then updated proportional to the prior times the weighted pseudo-likelihood, as in Equation (\ref{eq:pseudo_likelihood}), but using the draw-specific $\hat{w}_i^{(d)}$ and $\kappa^{(d)}$ values instead. 

Estimation of parameters proceeds via an MCMC Gibbs sampler. This is referred to as the \textit{fixed sampler}, where the model is run using the estimated number of patterns $\hat{K}$ from the adaptive sampler and the model parameter estimates are obtained. A post-processing variance adjustment is also used to ensure correct uncertainty estimation by the posterior samples \citep{williams2021uncertainty, wu2024derivation}. Without such a post-processing variance adjustment, posterior intervals will exhibit undercoverage as a consequence of model-misspecification that arises because the weighted pseudo-likelihood is a composite likelihood rather than a proper likelihood \citep{ribatet2012bayesian}. 

Finally, the posterior samples for $\mathbf{\pi}$ and $\mathbf{\theta}$ are stacked across the $D$ draws. The final estimates are obtained as the posterior median across all samples, and the posterior intervals are given by the quantiles across samples. These posterior estimates determine the composition of the latent class patterns. To prevent scenarios where the number of latent classes differs across the draws, we use a Dir($\hat{K}, \ldots, \hat{K}$) prior for $(\pi_1,\ldots, \pi_{\hat{K}})$, which puts higher probability on $\hat{K}$ latent classes being chosen, and we subset to the iterations that have a final number of latent classes equal to $\hat{K}$. Another complication of combining the MCMC posterior parameter samples across the D draws is alignment of class labels across draws. Each draw $d$ will have different class assignments $c_i$ that can correspond to differently estimated class profiles. To address this, we choose the class label permutation that results in the lowest mean absolute distance between the values of $\mathbf{\theta}$ across draws. We relabel the class assignments across the $D$ draws so that they are all in accordance, and then propagate this relabeling to the parameters. The final posterior median estimates for $\mathbf{\pi}$ and $\mathbf{\theta}$  are then used to update the latent class assignments, $c_i$, by sampling from its posterior distribution. This produces a set of class assignments, $c_i$, that can be used to recover classification in the original sample. Notably, this classification was not possible in previous papers that have applied multiple imputation approaches to mixture models \citep{gunawan2020bayesian}. 


\subsection{Association Analysis with Dependent Variables}
The output of the WOLCAN model also produces dietary behavior pattern assignments (i.e., latent class assignments), $c_i$, for all individuals in the NPS. We incorporate these assignments into a Bayesian survey-weighted logistic regression to estimate the association between the dietary behavior patterns and the outcomes of hypertension, type 2 diabetes, and hypercholesterolemia, generalized to the population of adults in Puerto Rico. For each of the $D$ draws of the estimated pseudo-weights, we apply a Bayesian survey-weighted logistic regression using a pseudo-likelihood approach with a post-processing variance adjustment. This accounts for possible informative sampling (i.e., when the outcome influences selection into the sample) and other selection biases, and allows for asymptotically correct point estimates and credible intervals with respect to the underlying population \citep{williams2021uncertainty}. Then, we combine posterior samples across the draws to obtain final estimates of the association between the dietary behavior pattern assignments and the outcomes of hypertension, diabetes, and hypercholesterolemia.

\section{Simulation Study}\label{sec:sims_WOLCAN}
We conduct two sets of simulation studies. Simulation set 1 evaluates the effectiveness of BART model prediction for estimating the pseudo-weights for the non-probability sample (Section \ref{subsec:weights_design}). Simulation set 2 evaluates the performance of the WOLCAN model in estimating true latent class profiles and prevalences in the population (Section \ref{subsec:lca_design}).

\subsection{Simulation Set 1: Pseudo-Weight Estimation}\label{subsec:weights_design}
\subsubsection{Pseudo-Weight Simulation Design}
We create a hypothetical population of size $N=40000$ with three auxiliary variables, $a_1$, $a_2$, and $a_3$, that influence selection into both the NPS $S_B$ and the PS $S_R$. Auxiliary variables $a_1$ and $a_2$ are generated from a multivariate Normal distribution with mean 0, variance 1, and covariance $\rho = 0.5$. Auxiliary variable $a_3$ is independently generated from a standard Normal distribution with mean 0 and variance 1.

\begin{figure}[!htb]
    \centering
    \includegraphics[width = 0.9\textwidth]{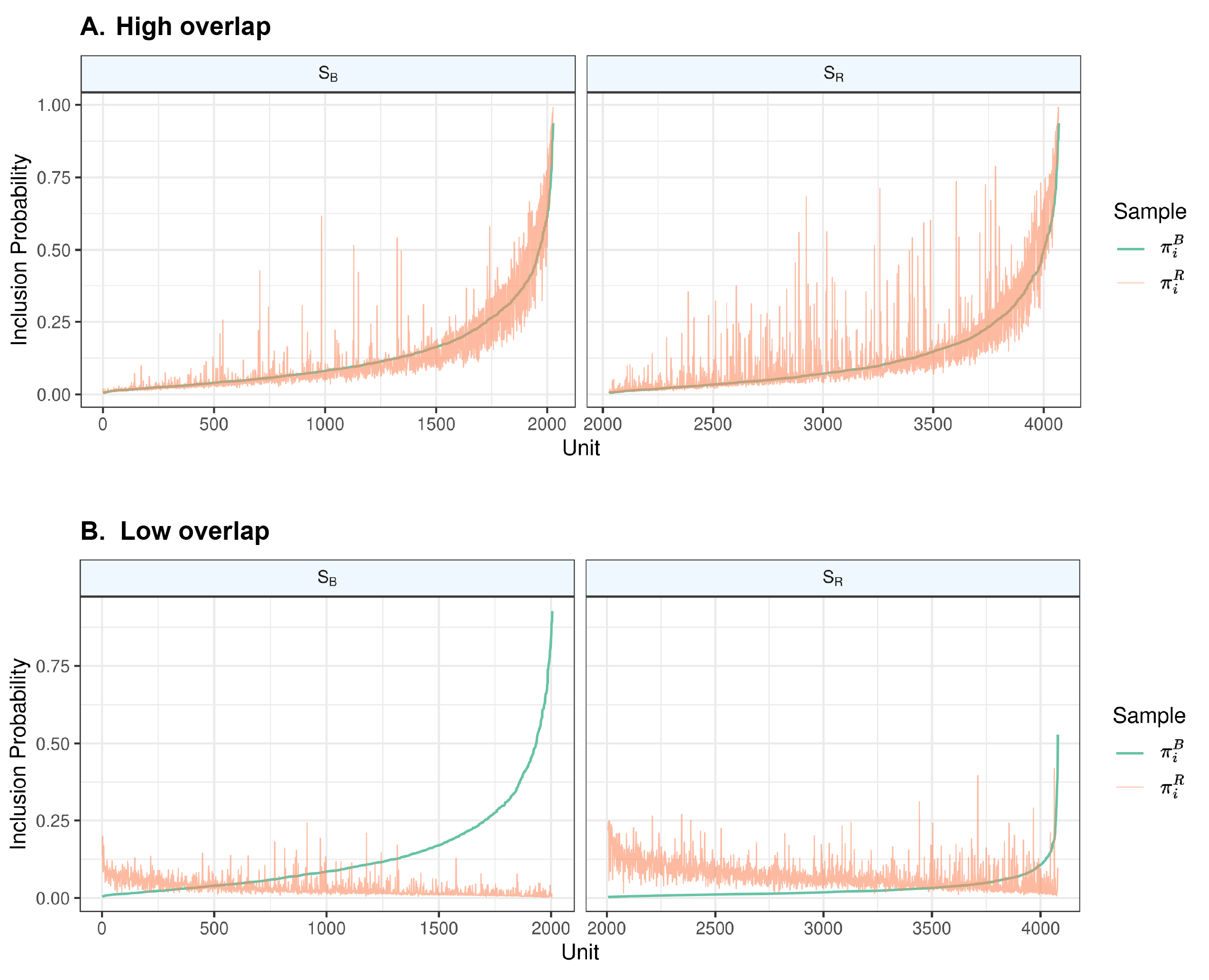}
    \caption{\textmd{Comparison of inclusion probabilities for the NPS, $\pi_i^B$, and inclusion probabilities for the PS, $\pi_i^R$, for units in $S_B$ and $S_R$ for a single realization, contrasting the high and low overlap settings.}}
    \label{fig:overlap_sims}
\end{figure}

An informative sampling design is adopted for both NPS and PS with unequal probabilities of inclusion that depend on the corresponding auxiliary variables. From this data, we simulate two scenarios: (1A) \textit{High overlap}, where the relationships between the auxiliary variables and the sample inclusion probabilities are similar for the NPS and PS (e.g., 7\% of individuals are shared across samples). (1B) \textit{Low overlap}, where the relationships between the auxiliary variables and sample inclusion are dissimilar for the NPS and PS (e.g., 2\% of individuals are shared across samples). Figure \ref{fig:overlap_sims} provides an example of a single realization's sample inclusion probabilities for the NPS and PS, respectively. The top row illustrates scenario 1A (high overlap). The bottom row illustrates scenario 1B (low overlap), where the inclusion mechanism differs greatly for the NPS and PS. 
For simplicity, clustering and stratification are not considered in the sampling design. We utilize a logistic function with complex interactions and non-linear terms.  For the high overlap setting, we have:
\begin{align}
    \pi_i^B &:= P(\delta_i^B = 1 | a_{i1}, a_{i2}, a_{i3})\nonumber\\
    &= \frac{\exp[\text{offset}_B - 0.9a_{i1} + 0.2a_{i1}^2 + 0.8 a_{i2} + 0.2 \log(a_{i2}) - 0.1 \sin{a_{i1}a_{i2}} + 0.3 a_{i3}]}{1+\exp[\text{offset}_B - 0.9a_{i1} + 0.2a_{i1}^2 + 0.8 a_{i2} + 0.2 \log(a_{i2}) - 0.1 \sin{a_{i1}a_{i2}} + 0.3 a_{i3}]}\label{eq:selection_probs_B}\\
    \pi_i^R &:= P(\delta_i^R = 1 | a_{i1}, a_{i2}, a_{i3})\nonumber\\
    &= \frac{\exp[\text{offset}_R - 0.6a_{i1} + 0.4 a_{i1}^2 + 0.7a_{i2} + 0.1\log(a_{i2}) - 0.05 \sin{a_{i1}a_{i2}} + 0.4 a_{i3}]}{1+\exp[\text{offset}_R - 0.6a_{i1} + 0.4 a_{i1}^2 + 0.7a_{i2} + 0.1\log(a_{i2}) - 0.05 \sin{a_{i1}a_{i2}} + 0.4 a_{i3}]}\label{eq:selection_probs_R},
\end{align}
where $\text{offset}_B$ and $\text{offset}_R$ denote offset adjustments for the intercepts so that the target sample sizes can be met. For the low overlap setting, we have more differences between the NPS and PS inclusion probability equations:
\begin{align}
    \pi_i^B &:= P(\delta_i^B = 1 | a_{i1}, a_{i2}, a_{i3})\nonumber\\
    &= \frac{\exp[\text{offset}_B - 0.9a_{i1} + 0.2a_{i1}^2 + 0.8 a_{i2} + 0.2 \log(a_{i2}) - 0.1 \sin{a_{i1}a_{i2}} + 0.3 a_{i3}]}{1+\exp[\text{offset}_B - 0.9a_{i1} + 0.2a_{i1}^2 + 0.8 a_{i2} + 0.2 \log(a_{i2}) - 0.1 \sin{a_{i1}a_{i2}} + 0.3 a_{i3}]}\label{eq:low_selection_probs_B}\\
    \pi_i^R &:= P(\delta_i^R = 1 | a_{i1}, a_{i2}, a_{i3})\nonumber\\
    &= \frac{\exp[\text{offset}_R + 0.7a_{i1} - 0.6 a_{i2} + 0.1\log(a_{i2}) + 0.1a_{i1}a_{i2} - 0.1 a_{i3}]}{1+\exp[\text{offset}_R + 0.7a_{i1} - 0.6 a_{i2} + 0.1\log(a_{i2}) + 0.1a_{i1}a_{i2} - 0.1 a_{i3}]}\label{eq:low_selection_probs_R}.
\end{align}

We draw 100 realizations each for the NPS and PS, using Poisson sampling with inclusion probabilities $P(\delta_i^B=1|\mathbf{a}_i)$ and $P(\delta_i^R=1|\mathbf{a}_i)$ defined by Equations (\ref{eq:selection_probs_B}) and (\ref{eq:selection_probs_R}) for high overlap and Equations (\ref{eq:low_selection_probs_B}) and (\ref{eq:low_selection_probs_R}) for low overlap. Four sample size settings are considered: 1) $n_B = n_R= 5\%$; 2) $n_B = 5\%$ and $n_R = 1\%$; 3) $n_B= 1\%, n_R= 5\%$; and 4) $n_B = n_R = 1\%$. Since Poisson sampling is used, sample sizes are approximate and may vary slightly among sample iterations, but will be $\sim$2000 for the 5\% settings and $\sim$400 for the 1\% settings.

We compare scenarios where $\pi_i^B$ and $\pi_i^R$ are estimated with BART (as described in Section \ref{subsec:pseudo_weights} ), with logistic regression and standard regression, commonly used among quasi-randomization approaches, as well as the reference ``no model" case where all weights are set to 1. To examine ability to recover complex and non-linear effects, the models to generate the pseudo-weights include covariates $a_1$, $a_2$, $a_3$, and $a_1:a_2$ interaction. For BART, we assess prediction performance when considering 500, 1000, and 2000 posterior samples retained. We also consider prediction performance when selection variable $a_3$ and the $a_1:a_2$ interaction term are omitted as covariates. We use R package \texttt{BART} \citep{chipman2010bart} to apply BART in the pseudo-weight estimation, with default values used for all parameters other than the number of posterior samples.

\subsubsection{Pseudo-Weight Simulation Results}

\begin{figure}
    \centering
    \includegraphics[width=\linewidth]{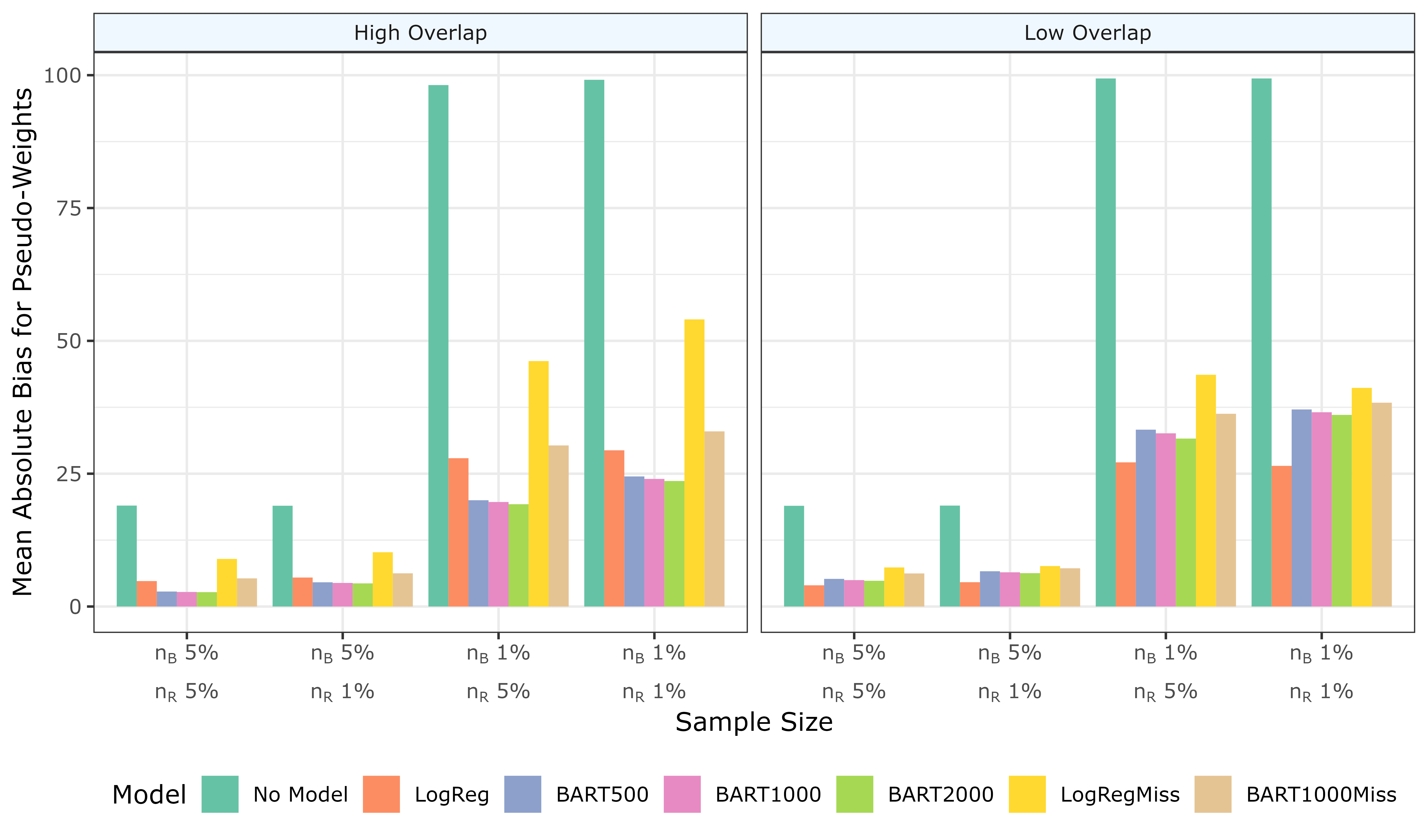}
    \caption{\textmd{Mean absolute bias of estimated pseudo-weights across different sample size scenarios and high and low covariate overlap scenarios. The models considered are: no model (all weights set to 1); logistic regression (LogReg); logistic regression with missing covariates (LogRegMiss); BART with 500, 1000, and 2000 posterior samples (BART500, BART1000, BART2000, respectively); BART with 1000 posterior samples and missing covariates (BART1000Miss)}}
    \label{fig:weights_sims}
\end{figure}

Figure \ref{fig:weights_sims} and Table \ref{table:weights_sims} display the mean absolute bias of the predicted pseudo-weights for the various models. Results are averaged across individuals and across the 100 random sample realizations, considering the high and low overlap settings and all sample size scenarios. Under the high overlap setting, the BART models uniformly outperform the logistic regression models when all covariates are included as well as when some covariates are missing. BART performance slightly improves with larger number of posterior samples but improvements are quite minor. Computationally, increasing the number of BART posterior samples does not result in large runtime differences, with runtimes for all three BART settings under one minute. Under the low overlap setting, logistic regression with all covariates performs well; however, when covariates are missing from the model, BART outperforms logistic regression and demonstrates better robustness. Small NPS sample size ($n_B \approx 400$) greatly increases the amount of bias in the predicted pseudo-weights, but this is to be expected given that individuals will have larger weights on average due to the smaller sample size. Overall, BART is more robust to missing covariates and interaction terms and has better performance under high overlap and large sample size settings. Based on the simulation results, a decision was made to consider a BART prediction model with 1000 posterior samples as the default model for the latent class simulation settings. 

\begin{table}[!htb]
\centering
\footnotesize
\caption{\textmd{Mean absolute bias of estimated pseudo-weights across different sample size scenarios and high and low covariate overlap scenarios.}}
\label{table:weights_sims}
\begin{tabular}{@{}lllllllll@{}}
\toprule
\textbf{Sample Size} & \textbf{Overlap} & \textbf{\begin{tabular}[t]{@{}l@{}}No\\ Model\end{tabular}} & \textbf{LogReg} & \textbf{\begin{tabular}[t]{@{}l@{}}LogReg\\ Miss\end{tabular}} & \textbf{\begin{tabular}[t]{@{}l@{}}BART\\ 500\end{tabular}} & \textbf{\begin{tabular}[t]{@{}l@{}}BART\\ 1000\end{tabular}} & \textbf{\begin{tabular}[t]{@{}l@{}}BART\\ 2000\end{tabular}} & \textbf{\begin{tabular}[t]{@{}l@{}}BART\\ 1000 Miss\end{tabular}} \\ \midrule
\multirow{2}{*}{$n_B$ 5\%, $n_R$ 5\%} & High & 19.00 & 4.79 & 8.96 & 2.82 & 2.74 & 2.71 & 5.30 \\
 & Low & 18.96 & 4.00 & 7.36 & 5.21 & 4.97 & 4.85 & 6.235 \\ \midrule
\multirow{2}{*}{$n_B$ 5\%, $n_R$ 1\%} & High & 18.97 & 5.46 & 10.22 & 4.56 & 4.44 & 4.36 & 6.25 \\
 & Low & 19.00 & 4.58 & 7.62 & 6.64 & 6.44 & 6.28 & 7.20 \\ \midrule
\multirow{2}{*}{$n_B$ 1\%, $n_R$ 5\%} & High & 98.14 & 27.91 & 46.20 & 20.01 & 19.67 & 19.26 & 30.32 \\
 & Low & 99.38 & 27.13 & 43.62 & 33.29 & 32.60 & 31.61 & 36.27 \\ \midrule
\multirow{2}{*}{$n_B$ 1\%, $n_R$ 1\%} & High & 99.13 & 29.40 & 54.04 & 24.49 & 24.02 & 23.62 & 32.97 \\
 & Low & 99.39 & 26.48 & 41.15 & 37.09 & 36.56 & 36.07 & 38.36 \\ \bottomrule
\end{tabular}
\end{table}

\subsection{Simulation Set 2: Latent Class Simulation} \label{subsec:lca_design}

\subsubsection{Latent Class Simulation Design}

For the latent class simulation analysis, we consider data generated under the following baseline setting. From a population of size $N=40000$, we draw 100 samples of sizes $n_R \approx 2000$ and $n_B \approx 2000$, each around 5\% of the population, using Poisson sampling with inclusion probabilities $P(\delta_i^B=1|\mathbf{a}_i)$ and $P(\delta_i^R=1|\mathbf{a}_i)$ defined by Equations (\ref{eq:selection_probs_B}) and (\ref{eq:selection_probs_R}) in the high overlap setting above. Coverage probabilities for the NPS and PS, $p_i^B$ and $p_i^R$, are set to 1. 

We generate the latent class data by simulating the latent class assignment variable, $c_i$, and the manifest categorical variables, $\mathbf{x}_i = (x_{i1},\ldots, x_{iJ})$ for all individuals $i$ in the population. We consider $K=3$ latent classes in the population, and we generate $c_i$ to depend on auxiliary variables $a_{i1}$, $a_{i2}$, and their interaction, $a_{i1}a_{i2}$, through a multinomial logistic regression formulation:
\begin{align}
    c_i|a_{i1}, a_{i2} &\sim \text{Multinomial}(1, \pi_{i1}, \ldots, \pi_{iK}), \\
    \quad \text{where } \eta_{ik} &:= \log\left(\frac{\pi_{ik}}{\pi_{i1}}\right) = \beta_{k0} + \beta_{k1}a_{i1} + \beta_{k2}a_{i2} + \beta_{k3}a_{i1}a_{i2},\\
    \text{with } \pi_{i1} &= \frac{1}{1+\sum\limits_{k=2}^K e^{\eta_{ik}}} \text{ and }
    \pi_{ik} = \frac{e^{\eta_{ik}}}{1+\sum\limits_{k=2}^K e^{\eta_{ik}}} \text{ for } k = 2,\ldots, K.
\end{align}
The following values are used for the $\beta$ parameters:
$$
\begin{pmatrix}\beta_{10} &\beta_{11} & \beta_{12} & \beta_{13} \\ \beta_{20} & \beta_{21} & \beta_{22} & \beta_{23} \\ \beta_{30}  & \beta_{31} & \beta_{32} & \beta_{33}  \end{pmatrix} = \begin{pmatrix}0 &0 & 0 & 0\\ 0.4 & -0.5 & 0.75 & 0.1\\ -0.2 & -1 & 1.2 & 0.25\end{pmatrix}. 
$$
The first row is set to $0$ for identifiability purposes. For $k=2$ and $k=3$, $\eta_{ik}$ models the dependence of the relative log-odds of $c_i=k$ compared to $c_i=1$ on the auxiliary variables. For example, $\exp(\hat{\beta}_{k0})$ is the odds of $c_i=k$ compared to $c_i=1$ when $a_{i1}=a_{i2}=0$, and $\exp(\hat{\beta}_{k1})$ is the odds ratio of $c_i=k$ compared to $c_i=1$ for a 1-unit increase in $a_{i1}$, holding the other model covariates fixed.

Next, we generate the multivariate categorical variables $\mathbf{x}_i = (x_{i1},\ldots, x_{iJ})$, with $J=30$ and $R=4$ category levels, to depend on $c_i$, $a_{i1}$, and $a_{i3}$ through $J$ multinomial logistic regression formulations. For each item $j$, define the item level probability for level $r$ to be $\theta_{ijr} := P(x_{ij}=r|c_i)$, with the constraints that $\theta_{ijr}>0$ and $\sum_{r=1}^R \theta_{ijr} = 1$. The latent class formulation sets $\theta_{ijr}$ equal to the class-specific item level probabilities, $\theta_{jkr} := P(x_{ij}=r|c_i=k)$, depending on whichever class $k$ is the observed value for $c_i$, so that those with similar $x_{ij}$ values are assigned to the same class. Since we have $K=3$, the corresponding multinomial logistic regression formulation for each item $j$ is given by:
\begin{align}
    x_{ij}|c_i, a_{i1}, a_{i3} &\sim \text{Multinomial}(1, \theta_{ij1},\ldots, \theta_{ijR})\\
    \text{where } \eta_{ijr} &:= \log\bigg(\frac{\theta_{ijr}}{\theta_{ij1}}\bigg) = \beta_{jr0} + \beta_{jr1}I(c_i=2) + \beta_{jr2}I(c_i=3) + \beta_{jr3}a_{i1} + \beta_{jr4}a_{i3} \nonumber \\
    &\qquad + \beta_{jr5}I(c_i=2)a_{i1} + \beta_{jr6}I(c_i=3)a_{i1} \text{ for } r=1,\ldots, R, \\
    \text{with } \theta_{ij1} &= \frac{1}{1+\sum\limits_{r=2}^R e^{\eta_{ijr}}}\text{ and }\
    \theta_{ijr} = \frac{e^{\eta_{ijr}}}{1+\sum\limits_{r=2}^R e^{\eta_{ijr}}} \text{ for } r= 2,\ldots, R. 
\end{align}
Note that this data generation process assumes a true latent class assignment variable $c_i$ exists without measurement error and determines the observed manifest variables $\mathbf{x}_i$, rather than the other way around.

\begin{figure}
    \centering
    \includegraphics[width = 0.6\textwidth]{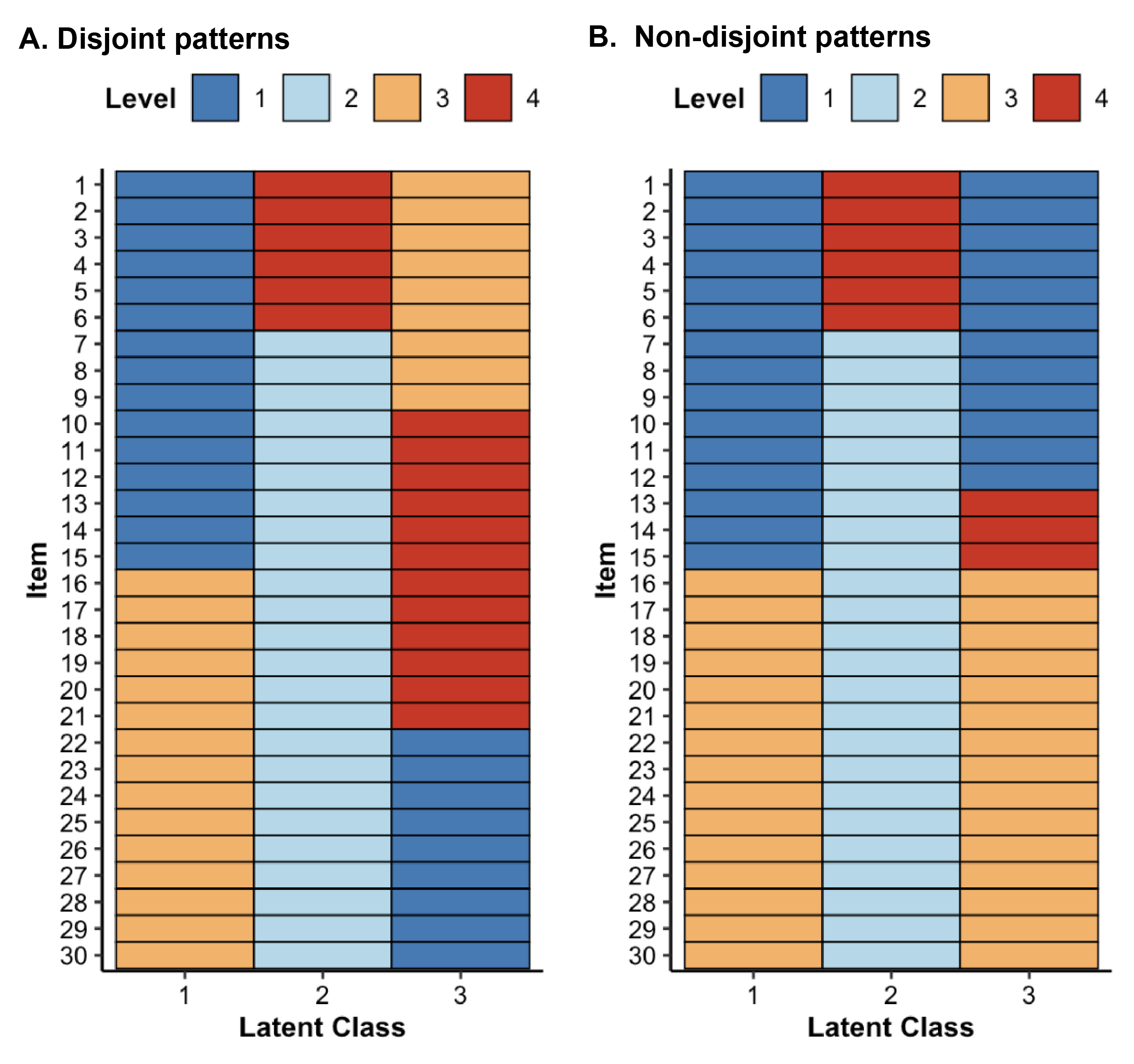}
    \caption{\textmd{Simulated modal item categories for the baseline setting with disjoint patterns (A) and for the setting with non-disjoint patterns where many items share the same modal category for two patterns (B). \textmd{Modal item category is defined as $\text{argmax}_r\theta_{jkr}$ for $r=1,\ldots,4,\ j = 1,\ldots, 30,\ k = 1,\ldots, 3$.}}}
    \label{f:sim_theta}
\end{figure}

The values for $\mathbf{\beta}_{jr} = (\beta_{jr0}, \beta_{jr1}, \ldots, \beta_{jr6})^\intercal$ are set so that $x_{ij}$ follows a Multinomial distribution where the modal (i.e., highest probability) level has probability around 0.85, and the remaining three levels have probability around 0.05. The modal levels have the following disjoint patterns for the $K=3$ latent classes: Pattern 1 has $r=1$ for the first 15 items and $r=3$ for the last 15; Pattern 2 has $r=4$ for the first 6 items and $r=2$ for the remaining 24; and Pattern 3 has $r=3$ for the first 9 items, $r=4$ for the next 12, and $r=1$ for the remaining 9. These modal pattern levels are displayed in Figure \ref{f:sim_theta}(a). The $\mathbf{\beta}_{jr}$ values are also chosen to induce association between $x_{ijr}$ and auxiliary variables influencing selection $a_{i1}$ and $a_{i3}$. For $j \in \{1,2\}$, $x_{ij}$ is associated with higher probability of $r=2$ across all classes. For $j\in \{29, 30\}$, $x_{ij}$ is associated with higher probability of $r=3$ for $k=1$, $r=2$ for $k=2$, and $r=2$ for $k=3$. The full list of values used for $\mathbf{\beta}_{jr}$, $j=1,\ldots, J$, $r=1,\ldots, R$, can be found in the Supplementary Materials. 

For the simulation study, the WOLCAN model is initialized with $K=30$ and sparsity-inducing prior $\text{Dir}(1/K, \ldots, 1/K)$ for the class membership probabilities $(\pi_1,\ldots, \pi_K)$. A noninformative flat $\text{Dir}(1)$ prior is used for the item level probabilities $(\theta_{jk1},\ldots, \theta_{jkR})$ for $j=1,\ldots, J$ and $k=1,\ldots, K$. Pseudo-weight estimation is run with covariates $a_1$, $a_2$, $a_3$, and $a_1:a_2$ interaction. Weights are trimmed using the IQR method with trimming constant $c=20$. To account for pseudo-weight uncertainty, model estimation proceeds by using $D=20$ draws from the posterior predictive distribution of the pseudo-weights.

We compare the performance of the proposed WOLCAN model with an unweighted Bayesian LCA model for various data generating scenarios beyond the described baseline. In addition to the baseline high overlap setting, we consider low overlap settings where the relationship between the covariates and selection greatly differ between the NPS and the PS, resulting in very few individuals being selected for both samples. The inclusion probabilities for low overlap are provided in Equations (\ref{eq:low_selection_probs_B}) and (\ref{eq:low_selection_probs_R}). We also consider non-disjoint latent class patterns, pictured in Figure \ref{f:sim_theta}(b), where patterns 1 and 3 are identical except for items 13-15. This examines the ability of the model to correctly separate out patterns that are similar except for a few influential items. Additionally, we simulate three different sample size settings: 5\% (baseline) with around 2000 individuals in each sample (NPS and PS); 1\% with around 400 individuals in each sample; and a sample size setting that closely follows the data application, where the NPS follows PROSPECT in including around 1500 individuals, the PS follows the PRCS in including around 75,000 individuals, and the population is set to 2,000,000 to approximate that of the current Puerto Rican adult population aged 30 to 75 years \citep{acs_2022_5year}. Finally, we assess model performance sensitivity when the prediction model for the weights is missing auxiliary variable $a_3$ and interaction terms. We also assess model performance when no post-processing variance adjustment is applied, as well as when only $D = 10$ draws from the pseudo-weight posterior are used to account for pseudo-weight estimation uncertainty. 

100 simulated datasets are generated for each scenario. To compare model performance for parameter estimation, we examine mean absolute bias (mean absolute distance between estimated and true parameter values), variability (full width of the 95\% credible interval (CI), averaged over latent classes), and coverage (proportion of 95\% CIs that cover the true population parameter values, averaged over latent classes).

\subsubsection{Latent Class Simulation Results}

\begin{figure}[!htb]
    \centering
    \includegraphics[width=\linewidth]{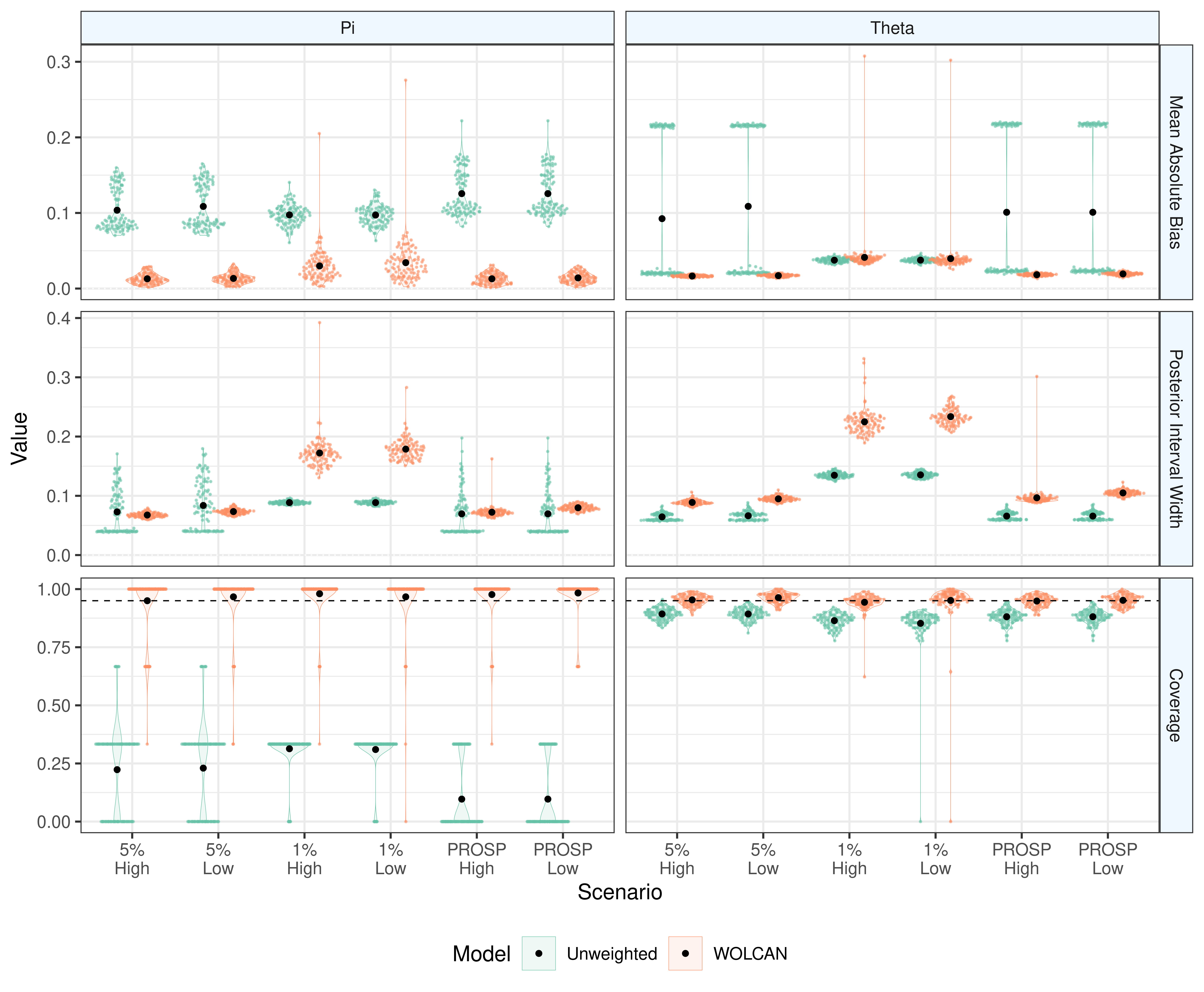}
    \caption{\textmd{Beeswarm and violin plot of mean absolute bias, 95\% interval width, and coverage for estimating class membership probabilities $\pi$ and item level probabilities $\theta$, for the unweighted model and the WOLCAN model, based on 100 independent samples. Results are displayed for scenarios considering 1\%, 5\%, and  PROSPECT-analogous sample sizes, and for scenarios considering high and low overlap between the NPS and PS. Dashed line indicates 0.95 coverage. Black dots indicate mean values across samples.}}
    \label{fig:scenario_sims}
\end{figure}

In Figure \ref{fig:scenario_sims}, beeswarm and violin plots show the model performance results for the unweighted model compared to the proposed WOLCAN model. The figure displays the mean absolute bias, 95\% posterior interval width, and coverage for the class membership probabilities, $\pi$, and class-specific item level probabilities, $\theta$, averaged all latent classes and across 100 sample iterations. For $\theta$, the metrics are calculated using the modal item level probabilities, $\text{argmax}_r\theta_{jkr}$.

Performance is compared across three sample sizes: 5\%, 1\%, and PROSPECT-analogous. The settings of high overlap and low overlap between the NPS and the PS are considered. For all sample size and overlap scenarios, the WOLCAN model is able to attain approximately nominal coverage of all parameters with minimal bias. In contrast, the unweighted model shows substantial bias, severe undercoverage, and issues with mis-estimation of pattern profiles and prevalence. In particular, estimation of $\pi$ is poor for all scenarios when using the unweighted model. The posterior interval width is mostly comparable between the two models, although WOLCAN displays more variability in the 1\% sample size setting. 

Table \ref{table:lca_results} displays results for all simulation scenarios and also includes results assessing bias in estimation of the pseudo-weights and the number of latent classes, $K$. We see that the unweighted model incorrectly estimates $K$ at a much higher rate, whereas the WOLCAN model is almost always able to estimate $K$ correctly. This mis-estimation of $K$ for the unweighted model also contributes to the higher bias in estimating the pattern profiles, $\theta$. 

\begin{figure}
    \centering
    \includegraphics[width=0.9\linewidth]{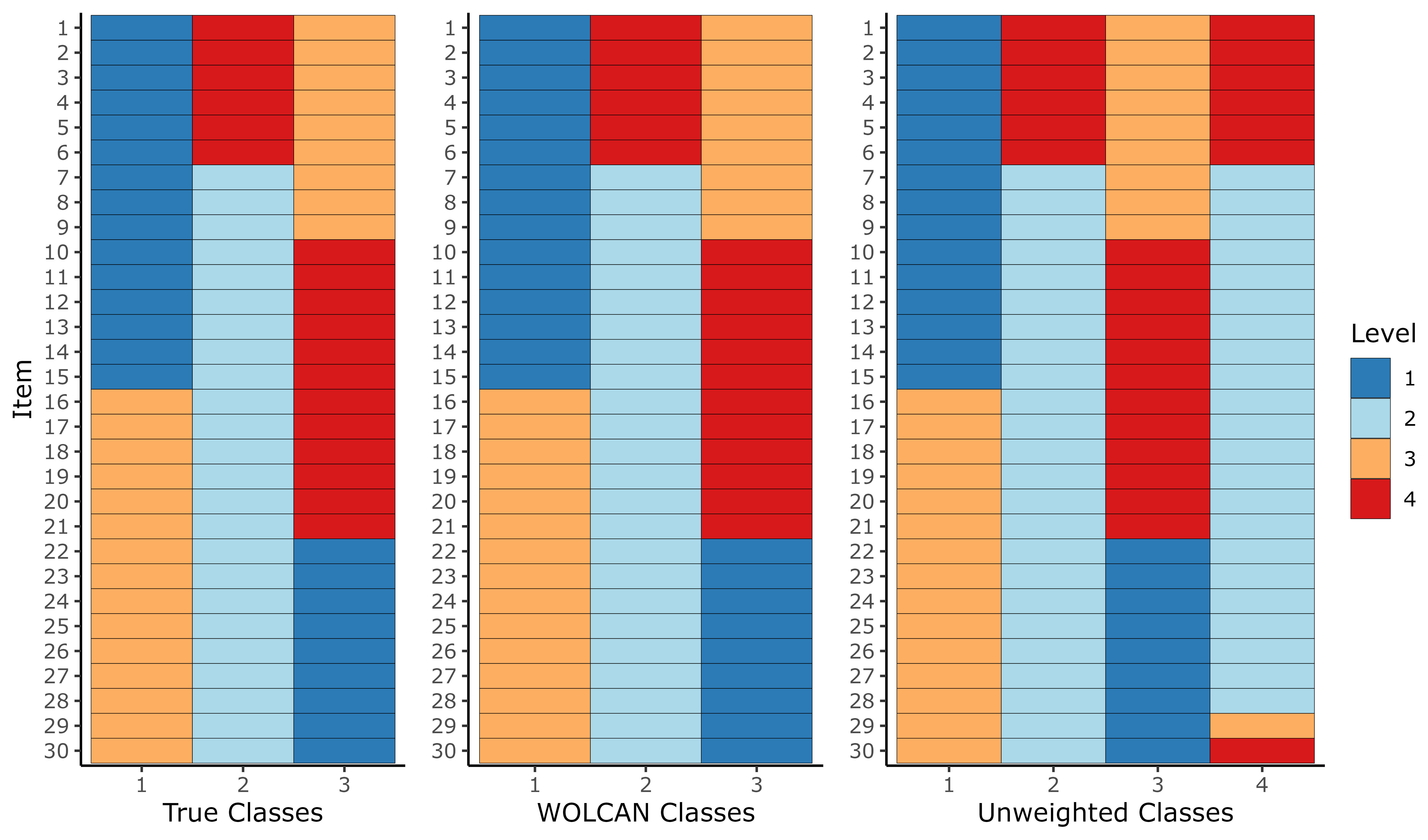}
    \caption{\textmd{Comparison of true (left) modal item response levels characterizing the latent class patterns with the estimated patterns from the WOLCAN model (center) and the unweighted model (right). Results are shown for a single sample realization.}}
    \label{fig:theta_modes}
\end{figure}

In the baseline setting (scenario 2A), the WOLCAN model is able to correctly identify the $K=3$ patterns for all 100 sample iterations, while the unweighted model is incorrect for 37 of those iterations and mistakenly identifies four classes. An example for a single iteration is provided in Figure \ref{fig:theta_modes}, which displays the true latent classes alongside the estimated classes using the WOLCAN and unweighted models. The unweighted model identifies a spurious fourth class that resembles the second latent class but has differing estimated modes for items 29 and 30. This is likely due to items 29 and 30 being generated to be directly associated with auxiliary variable $a_1$ that also influences selection into the samples. We see that when there are variables that are associated with selection and also directly influence item level probabilities beyond the latent class assignment probabilities, then mis-estimation of the true latent patterns can occur. This illustrates the importance of patterns accounting for the selection mechanism. 

Regarding WOLCAN estimation of the pseudo-weights, there is more bias when the sample size is very small. However, this is to be expected due to the larger values of the individual weights arising from the sample. The model still performs well in terms of estimation of $\pi$ and $\theta$. Notably, the model performs well in the PROSPECT-analogous setting where the NPS sample size is very small. These trends are also reflected in the sensitivity analyses in scenarios 2G, 2H, 2I, and 2J. For non-disjoint underlying patterns, the WOLCAN model performs well, though its advantages over the unweighted model are slightly attenuated due to the increased variability in the parameter estimates and the improved performance of the unweighted model in estimating the true number of latent classes. When only $D=10$, rather than $D=20$, draws are taken from the pseudo-weight posterior to account for pseudo-weight estimation uncertainty,  the WOLCAN model still shows comparable bias, variance, and coverage for all parameters. Computational runtimes were around 3.5 hours for $D=20$ and around 2.5 hours for $D=10$, after parallelizing over 8 cores. Good performance was also maintained under the setting where the pseudo-weight prediction model was missing selection variable $a_3$ and interaction terms. A slight increase in the pseudo-weight prediction error had minimal impact on parameter estimation. Coverage decreases when no post-processing variance adjustment is applied, but bias remains minimal. Overall, the WOLCAN model is able to estimate all model parameters with minimal bias while obtaining nominal coverage of posterior intervals, whereas the unweighted model results in bias and undercoverage. 

\begin{table}
\caption{\textmd{Summary of performance metrics for the WOLCAN and unweighted models over 100 independent samples. Abbreviations: wts = weights, abs = absolute, CI = credible interval, cov = coverage. Scenarios: 2A = sample size (SS) 5\% high overlap (baseline), 2B = SS 5\% low overlap, 2C = SS 1\% high overlap, 2D = SS 1\% low overlap, 2E = SS PROSPECT high overlap, 2F = SS PROSPECT low overlap, 2G = baseline with non-disjoint patterns, 2H = baseline with D = 10, 2I = baseline with missing selection covariates, 2J = baseline with no variance adjustment.}}
\scriptsize
\resizebox{\textwidth}{!}{%
\begin{tabular}{@{}llllllllll@{}}
\toprule
\textbf{Scenario} & \textbf{Model} & \textbf{\begin{tabular}[t]{@{}l@{}}Wts Abs \\ Bias\end{tabular}} & \textbf{\begin{tabular}[t]{@{}l@{}}$K$ Abs \\ Bias\end{tabular}} & \textbf{\begin{tabular}[t]{@{}l@{}}$\pi$ Abs \\ Bias\end{tabular}} & \textbf{\begin{tabular}[t]{@{}l@{}}$\theta$ Abs \\ Bias\end{tabular}} & \textbf{\begin{tabular}[t]{@{}l@{}}$\pi$ CI \\ Width\end{tabular}} & \textbf{\begin{tabular}[t]{@{}l@{}}$\theta$ CI \\ Width\end{tabular}} & \textbf{\begin{tabular}[t]{@{}l@{}}$\pi$ \\ Cov\end{tabular}} & \textbf{\begin{tabular}[t]{@{}l@{}}$\theta$ \\ Cov\end{tabular}} \\ \midrule
\multirow{2}{*}{2A} & Unweighted & 19.00 & 0.37 & 0.104 & 0.034 & 0.073 & 0.064 & 0.223 & 0.893 \\
 & WOLCAN & 2.74 & 0.00 & 0.013 & 0.012 & 0.068 & 0.089 & 0.950 & 0.954 \\ \midrule
\multirow{2}{*}{2B} & Unweighted & 18.96 & 0.45 & 0.109 & 0.039 & 0.084 & 0.066 & 0.230 & 0.893 \\
 & WOLCAN & 4.97 & 0.00 & 0.013 & 0.012 & 0.073 & 0.095 & 0.967 & 0.963 \\ \midrule
\multirow{2}{*}{2C} & Unweighted & 99.13 & 0.00 & 0.097 & 0.024 & 0.089 & 0.135 & 0.313 & 0.864 \\
 & WOLCAN & 24.02 & 0.01 & 0.030 & 0.028 & 0.172 & 0.225 & 0.980 & 0.943 \\ \midrule
\multirow{2}{*}{2D} & Unweighted & 99.28 & 0.00 & 0.097 & 0.024 & 0.089 & 0.135 & 0.310 & 0.853 \\
 & WOLCAN & 36.48 & 0.01 & 0.034 & 0.026 & 0.179 & 0.234 & 0.967 & 0.951 \\ \midrule
\multirow{2}{*}{2E} & Unweighted & 1995.97 & 0.07 & 0.106 & 0.019 & 0.056 & 0.071 & 0.203 & 0.876 \\
 & WOLCAN & 282.04 & 0.00 & 0.015 & 0.015 & 0.083 & 0.110 & 0.977 & 0.945 \\ \midrule
\multirow{2}{*}{2F} & Unweighted & 1995.97 & 0.07 & 0.106 & 0.019 & 0.056 & 0.071 & 0.203 & 0.876 \\
 & WOLCAN & 588.41 & 0.00 & 0.017 & 0.016 & 0.094 & 0.125 & 0.973 & 0.949 \\ \midrule
\multirow{2}{*}{2G} & Unweighted & 18.99 & 0.03 & 0.090 & 0.019 & 0.049 & 0.061 & 0.307 & 0.848 \\
 & WOLCAN & 2.73 & 0.00 & 0.012 & 0.012 & 0.093 & 0.098 & 0.957 & 0.941 \\ \midrule
\multirow{2}{*}{2H} & Unweighted & 19.00 & 0.37 & 0.104 & 0.034 & 0.073 & 0.064 & 0.223 & 0.893 \\
 & WOLCAN & 2.74 & 0.00 & 0.013 & 0.012 & 0.069 & 0.092 & 0.953 & 0.954 \\
 \midrule
\multirow{2}{*}{2I} & Unweighted & 19.00 & 0.37 & 0.104 & 0.034 & 0.073 & 0.064 & 0.223 & 0.893 \\
 & WOLCAN & 3.06 & 0.00 & 0.013 & 0.012 & 0.069 & 0.090 & 0.943 & 0.953 \\ \midrule
 \multirow{2}{*}{2J} & Unweighted & 19.00 & 0.37 & 0.104 & 0.034 & 0.073 & 0.064 & 0.223 & 0.893 \\
 & WOLCAN & 2.74 & 0.00 & 0.013 & 0.012 & 0.047 & 0.060 & 0.870 & 0.842 \\ \bottomrule
\end{tabular}%
}
\label{table:lca_results}
\end{table}

\section{Application to Derive Dietary Behavior Patterns in Puerto Rico}\label{sec:application}
We illustrate the utility of our model in eliciting dietary behavior patterns representative of the population of adults living in Puerto Rico and measuring their associations with cardiometabolic outcomes. We define our reference probability sample using data from the Puerto Rico Community Survey (PRCS), which will provide sample selection information through a known survey design. We define our non-probability sample using data from the Puerto Rico Observational Study of Psychosocial, Environmental, and Chronic Disease Trends (PROSPECT), which includes the exposure and outcome variables of interest. We estimate pseudo-weights for this NPS and then conduct a dietary behavior pattern analysis.

\subsection{Puerto Rico Community Survey}

Conducted by the U.S. Census Bureau and belonging to the American Community Survey (ACS), the PRCS provides extensive demographic, housing, social, and economic data for communities throughout Puerto Rico on a yearly basis and has done so since 2005 \citep{bureau2020understanding}. Independent housing units are sampled for each of the municipalities in PR, averaging about 36,000 housing units sampled each year. A subset of individual-level data is publicly available through public use microdata sample (PUMS) files that have been stripped of personally identifiable information for confidentiality purposes \citep{bureau2021understanding}. These data are available through five-year estimates, which aggregate data from five consecutive years to produce reliable estimates for all regions in Puerto Rico, including lesser populated areas where one-year estimates are not available. We use the most recent five-year estimates from 2018 to 2022. The sampling design includes multi-stage stratified cluster sampling, conducted separately for housing units and group quarters. To enable representative estimates and standard errors, sampling weights are provided that account for the complex survey design elements and incorporate post-stratification, raking, and non-response adjustments \citep{bureau2022sample}. 

The pseudo-weight generation model incorporated the following PRCS auxiliary variables to characterize the selection mechanism: age, sex, educational attainment, annual household income, and ethnicity. These variables were chosen for their availability in both the PRCS and PROSPECT datasets. Individual-level urbanicity was excluded as it was not available in the PRCS; however, because BART was used to generate pseudo-weights, the approach may remain robust despite this missing variable, as demonstrated in the simulations. Individuals were restricted to the ages 30 to 75 to match the age restriction of PROSPECT. A total of $n_R=77,907$ individuals with complete data were included in the analysis. Assuming a population of around 2 million adults of that age living in PR, this corresponds to a sampling percentage of around 3.9\%.

\subsection{Puerto Rico Observational Study of Psychosocial, Environmental, and Chronic Disease Trends} 
Our NPS dataset comes from the Puerto Rico Observational Study of Psychosocial, Environmental, and Chronic Disease Trends (PROSPECT). PROSPECT is an observational cohort study that aims to identify trends in socioeconomic, environmental, behavioral, psychosocial, and biological factors associated with cardiovascular disease among adults aged 30 to 75 across all of PR \citep{mattei2021design}. Baseline data are available for around 1,500 adults, collected during 2019-2024 at partner clinic sites with trained, bilingual staff. Eligibility criteria include 30 to 75 years of age, residence in PR at least 12 months, ability to respond to questions without assistance, and non-institutionalized and non-transient housing. Participants were recruited using a combination of various community-based sampling techniques such as recruitment from health fairs and clinics, social media, and professional workers unions, and with some individuals recruited through the original multistage, probabilistic sampling strategy. A total of $n_B=1550$ individuals with complete data were included in this analysis. Assuming a population of around 2 million adults living in PR, this corresponds to a sampling percentage of around 0.08\%. 

The selection variables of age, sex, educational attainment, annual household income, and ethnicity were used to model the underlying selection mechanism into the study. Selection into PROSPECT is assumed to depend only on these variables, but this dependence can be non-linear and include interactions. Harmonization of selection variable measurement was used to ensure consistent measurement across both PRCS and PROSPECT. 

Dietary behavior exposure variables were self-reported as part of a dietary behaviors questionnaire assessing purchasing and cooking roles, eating frequency and timing, eating out behavior, eating practices, dieting practices, weight and supplements, nutrition awareness and knowledge, and sustainability practices. A total of 35 variables were used, each treated as a two- or three-level categorical variable, with potential recoding to increase ordering congruency and minimize small cell sizes. Three-level variables included categories for Low, Medium, or High risk associations with cardiometabolic conditions, and two-level variables included categories for Low or Medium risk. Table \ref{tab:dietary_behavior} displays the assumed ordering of categories assigned to the different risk levels for each dietary behavior variable.

Three binary cardiometabolic outcomes were considered: hypertension, type 2 diabetes, and hypercholesterolemia. Hypertension was defined as having either self-reported diagnosis, self-reported use of hypertension-controlling medication, and/or an elevated blood pressure (BP) reading (systolic BP $>$ 130 or diastolic BP $>$ 80; \citet{whelton20182017}). Type 2 diabetes was defined as having self-reported diagnosis, self-reported use of diabetes-controlling medication, and/or a laboratory-based measure (fasting blood sugar $\geq$ 126 mg/dL or hemoglobin A1c $\geq$ 6.5; \citet{ada2021diabetes}). Hypercholesterolemia was defined as having self-reported diagnosis, self-reported use of cholesterol-controlling medication, and/or a laboratory-based total cholesterol measure $>$ 200 mg/dL \citep{grundy20192018}.

Additional self-reported sociodemographic and behavioral variables were incorporated to describe the distribution of the derived dietary behavior patterns and as potential confounders for the association between dietary behavior patterns and the outcomes of interest. The following variables were considered: age; sex at birth; educational attainment (less than high school, high school, some college or bachelor's degree, and some graduate education); annual household income ($\leq$ \$10,000; \$10,001 to \$20,000; $>$\$20,000); ethnicity (Puerto Rican or other); urbanicity (urban area of residence or rural/in-between), smoking status (never, former, or current); drinking status (never, former, or current); physical activity (active or sedentary); food security (secure or insecure); receipt of benefits from the federal supplemental nutrition program for women, infants, and children (WIC) and/or the Puerto Rican nutrition assistance program (PAN); social support; perceived stress; depressive symptoms (none/mild or moderate/severe); and generalized anxiety (none/mild or moderate/severe). Additional details on all variables can be found in the Supplementary Materials, with further data collection details and methodology provided elsewhere \citep{mattei2021design}. 

The WOLCAN model was initialized with the same priors used in the simulation study. The pseudo-weights were generated using BART with 1000 posterior samples, and weight trimming was applied to weights that exceeded the median by 20 times the inter-quartile range. Estimation of the dietary behavior patterns was obtained by fitting a Gibbs sampler of 20,000 iterations with 10,000 burn-in and thinning every 5 iterations. Parameters were summarized using posterior median estimates. Variance adjustment for pattern estimation was not performed due to lack of smoothness in the posterior, but analyses remain valid because only point estimates were used in characterizing the pattern profiles. Variance adjustment was performed for the Bayesian survey-weighted outcome regression. Computation time for the WOLCAN model was roughly 4 hours using an Apple M1 Pro computer with 8 cores. To avoid potential reverse causation where dietary behavior is modified due to known diagnosis, a sensitivity analysis was performed on the outcome models to evaluate if results differed when self-reported outcome diagnoses were removed from the analysis. Each outcome model had a runtime of approximately 20 minutes. All analyses were conducted using the R software environment (version 4.3.1) and R packages (non-exhaustive) \texttt{BART} \citep{chipman2010bart}, \texttt{baysc} \citep{wu2025baysc}, \texttt{csSampling} \citep{hornby2023cssampling}, \texttt{brms} \citep{burkner2017brms}, and \texttt{survey} \citep{lumley2024survey}. 

\begin{singlespacing}
\DefTblrTemplate{middlehead,lasthead}{default}{}
\begin{longtblr}[
caption = {Description and coding levels for the 35 dietary behavior variables included. Response levels were recoded so that health-related behaviors were low risk and adverse behaviors were high risk.},
  label = {tab:dietary_behavior}
                    ]{colsep  = 3pt,
                    colspec = {l X[0.9, l] X[3, l] X[2.7,l]},
                      rows    = {font=\scriptsize},
                      row{1}  = {font=\scriptsize\bfseries},
                      rowsep  = 0.5pt,
                      rowhead = 1,
                      }
    \toprule
& Variable & Description & Response Risk Levels  \\
  \midrule
\SetCell[c=4]{l}{\bfseries Purchasing and Cooking Roles}\\ \hline
1. &Food Purchase Person & Who is the person that usually purchases the food that you cook and consume in your household? & 
Low = Myself (participant) \newline
Med = Partner, family member, or other people \newline
High = I usually do not eat at my house \\ \hline
2. &Food Purchase Location & Where do you buy the majority of the food that you cook? & 
Low = Supermarket chain (e.g., Pueblo, Amigo, Econo) \newline
Med = Discount stores with food like Walmart and Costco \newline
High = Convenience store in your neighborhood, or other \\ \hline
3. &Food Purchase Frequency & How many times do you enter any type of establishments to buy food (full grocery shopping); not only to buy two or three articles, like milk or bread? & 
Low = Once a week or more \newline
Med = Once every two weeks \newline
High = Once a month or less \\ \hline
4. &Cooking Person & Who is the person that usually prepares food/cooks the meals that you consume in your household? & 
Low = Myself (participant) \newline
Med = Partner, family member, or other people \newline
High = I usually do not eat at my house \\ \hline
\SetCell[c=4]{l}{\bfseries Eating Frequency}\\ \hline
5. &Breakfast Frequency & How often do you eat a full breakfast, and not only coffee and a “bite of something” (like a donut or toast)? & 
Low = Many times (5-7 times per week) \newline
Med = Sometimes (2-4 times per week) \newline
High = Rarely or never \\ \hline
6. & Lunch Frequency & How many times do you eat a full lunch? & 
Low = Many times (5-7 times per week) \newline
Med = Sometimes (2-4 times per week) \newline
High = Rarely or never \\ \hline
7. & Dinner Frequency & How many times do you eat a full dinner? & 
Low = Many times (5-7 times per week) \newline
Med = Sometimes (2-4 times per week) \newline
High = Rarely or never \\ \hline
8. & Snack Frequency & How many times do you eat snacks? & 
Low = Rarely or never \newline
Med = Sometimes (2-4 times per week) \newline
High = Many times (5-7 times per week) \\ \hline
\SetCell[c=4]{l}{\bfseries Eating Out}\\ \hline
9. & Fast Food Frequency & On average, on a usual week, how many times do you eat commercial meals prepared outside your household in fast food establishments? & 
Low = Rarely or never \newline
Med = Sometimes (2-4 times per week) \newline
High = Many times (5-7 times per week) \\ \hline
10. & Restaurant Frequency & On average, on a usual week, how many times do you eat commercial meals prepared outside your household in restaurants? & 
Low = Rarely or never \newline
Med = Sometimes (2-4 times per week) \newline
High = Many times (5-7 times per week) \\ \hline
11. & Take Out Frequency & On average, on a usual week, how many times do you eat commercial meals prepared outside your household as take out? & 
Low = Rarely or never \newline
Med = Sometimes (2-4 times per week) \newline
High = Many times (5-7 times per week) \\ \hline
12. & Food Truck Frequency & On average, on a usual week, how many times do you eat commercial meals prepared outside your household in food trucks? & 
Low = Rarely or never \newline
Med = Sometimes (2-4 times per week) \newline
High = Many times (5-7 times per week) \\ \hline
13. & Café and Bakery Frequency & On average, on a usual week, how many times do you eat commercial meals prepared outside your household in cafés or bakeries? & 
Low = Rarely or never \newline
Med = Sometimes (2-4 times per week) \newline
High = Many times (5-7 times per week)  \\ \hline
14. & Party Frequency & On average, how many times per week do you attend activities, parties or celebrations outside your household where food is served? & 
Low = Rarely or never \newline
Med = Sometimes (2-4 times per week) \newline
High = Many times (5-7 times per week)  \\ \hline
\SetCell[c=4]{l}{\bfseries Eating Practices}\\ \hline
15. & Meal Alone Frequency & On a usual week, how often do you eat a meal (lunch, breakfast or dinner) by yourself (in your car, at work, at home but alone)? & 
Low = Rarely or never \newline
Med = Sometimes (2-4 times per week) \newline
High = Many times (5-7 times per week) \\ \hline
16. & Meal TV Frequency & On a usual week, how often do you eat a meal (breakfast, lunch or dinner) while you watch TV? & 
Low = Rarely or never \newline
Med = Sometimes (2-4 times per week) \newline
High = Many times (5-7 times per week) \\ \hline
17. & Meal Quality & In general, how would you describe your current eating habits and the quality of your meals? & 
Low = Excellent or very good \newline
Med = Good \newline
High = Fair or poor  \\ \hline
18. & Meal Healthy & Nowadays, how often do you modify or change your meals to make them healthier? & 
Low = Many times (5-7 times per week) \newline
Med = Sometimes (2-4 times per week)  \newline
High = Rarely or never  \\ \hline
19. & Meal Acculturation & In general, between typical American or typical Puerto Rican meals, which type do you consume the most? & 
Low = Mostly Puerto Rican meals \newline
Med = Same amount of Puerto Rican and American meals \newline
High = Mostly American meals  \\ \hline
\SetCell[c=4]{l}{\bfseries Dieting Practices}\\ \hline
20. & Control Salt & How often do you control salt while cooking or eating? & 
Low = Many times \newline
Med = Sometimes \newline
High = Rarely or never \\ \hline
21. & Control Fat & How often do you control fat while cooking or eating? & 
Low = Many times \newline
Med = Sometimes \newline
High = Rarely or never \\ \hline
22. & Control Carbs and Sugars & How often do you control carbs and sugars while cooking or eating? & 
Low = Many times \newline
Med = Sometimes \newline
High = Rarely or never \\ \hline
23. & Control Portions & How often do you control portions while cooking or eating? & 
Low = Many times \newline
Med = Sometimes \newline
High = Rarely or never \\ \hline
24. & Count Calories & How often do you count calories while cooking or eating? & 
Low = Many times \newline
Med = Sometimes \newline
High = Rarely or never \\ \hline
\SetCell[c=4]{l}{\bfseries Weight and Supplements}\\ \hline
25. & Dysfunctional Weight Loss & Do you do any of the following to lose weight: ipecac, laxatives, throw up/vomit, or take diuretics? & 
Low = No \newline
Med = Yes \\ \hline
26. & Vitamins & During the last month, have you used vitamins or other dietary supplements? & 
Low = Once or twice a day \newline
Med = 2-3 days per week \newline
High = No \\ \hline
27. & Probiotics & During the past 3 months, have you consumed natural probiotics (yogurts or drinks enriched with probiotics) or commercial probiotic supplements? & 
Low = Yes, once or twice a day \newline
Med = Yes, 3 days per week or less \newline
High = No \\ \hline
28. & Check Weight & How often do you check your body weight? & 
Low = Everyday \newline
Med = 1 to 4 times per month \newline
High = Once per month or less  \\ \hline
\SetCell[c=4]{l}{\bfseries Nutrition Awareness and Knowledge}\\ \hline
29. & Food Guide & Have you ever seen the “My Plate” food guide? & 
Low = Yes \newline
Med = No  \\ \hline
30. & Nutrition Panel Frequency & How often do you read the nutrition fact panel (label with nutritional information) on a food product? & 
Low = Every day or nearly everyday \newline
Med = A few times per week \newline
High = A few times per month or less \\ \hline
31. & Nutrition Info Frequency & How often do you look for or use information about eating healthy and nutrition extracted from TV, magazines, internet, and social media? & 
Low = Every day or nearly everyday \newline
Med = A few times per week \newline
High = A few times per month or less \\ \hline
\SetCell[c=4]{l}{\bfseries Sustainability}\\ \hline
32. & Local Food Frequency & How often do you purposely purchase foods from Puerto Rico (like fresh fruits, vegetables, meat and other products that are produced here rather than being imported)? & 
Low = Many times (5-7 times per week) \newline
Med = Sometimes (2-4 times per week) \newline
High = Rarely or never \\ \hline
33. & Buy Organic Frequency & How often do you buy organic food or products for cooking or eating? & 
Low = Many times (5-7 times per week) \newline
Med = Sometimes (2-4 times per week) \newline
High = Rarely or never  \\ \hline
34. & Water Type & Which of these types of drinking water do you use at home most of the time? & 
Low = Right from the faucet without filtering \newline
Med = Filtered or boiled water from the faucet \newline
High = Bottle water or other \\ \hline
35. & Eat Vegetarian & How often do you eat vegetarian or less meats while cooking or eating? & 
Low = Many times \newline
Med = Sometimes \newline
High = Rarely or never \\ \hline
\end{longtblr}
\end{singlespacing}

\subsection{Results}

\subsubsection{Pseudo-Weight Estimation}
The distribution of estimated pseudo-weights was unimodal with considerable right skew, indicating that a few individuals from under-sampled groups represent a relatively large proportion of the population. The posterior mean weights had a median of 603.9, a maximum of 17114.5, and an interquartile range of 826.6. Table \ref{tab:weights_res} provides a comparison of how the PROSPECT sample and the target population differ in terms of the distribution of the selection variables, as well as other sociodemographic and behavioral variables. In particular, it displays the distributions of the variables among those in the PROSPECT sample, those in the sample above the median weight, those with very high weight, and those in the target population. Population distributions were estimated using the pseudo-weights. For the selection variables with data available from the reference PRCS sample, validation checks showed close alignment of the population distributions, indicating reliable performance of the pseudo-weight predictions (Appendix Table 1). 

Among the variables incorporated in the selection model, individuals who were male, had high school or lower educational attainment, and/or were higher income were under-represented in the PROSPECT sample. These individuals were given higher weights in order to more closely align the weighted sample to the population composition. Among variables that were not used in the selection model, there were also certain subgroups that were under-represented in the sample and up-weighted, though differences were comparatively more minor. These subgroups included individuals who are current smokers, former or current drinkers, physically active, food secure, non-WIC/PAN recipients, and/or with none or mild depressive symptoms. Conversely, sample and population composition were relatively balanced for variables relating to age, ethnicity, social support, perceived stress, and anxiety. 

\begin{table}[!htb]
\caption{\textmd{Distribution of selection variables and other sociodemographic and behavioral variables among all individuals in the PROSPECT sample, among those with estimated weights greater than the median (Q50), among those with estimated weights greater than the 90-th percentile (Q90), and among all those in the target population. Population values are estimated by upweighting sampled individuals according to their estimated weights. Means are reported for continuous variables, and column percentages are reported for categorical variables.}}
\scriptsize
\centering
\begin{tabular}{@{}llcccc@{}}
\toprule
\begin{tabular}[c]{@{}c@{}}Variable\\ \cr \end{tabular} & \begin{tabular}[c]{@{}c@{}}Level\\ \cr \end{tabular} & \begin{tabular}[c]{@{}c@{}}PROSPECT\\ Sample \end{tabular} & \begin{tabular}[c]{@{}c@{}}Weights\\ $>$Q50 \end{tabular} & \begin{tabular}[c]{@{}c@{}}Weights\\ $>$Q90 \end{tabular} & \begin{tabular}[c]{@{}c@{}}Population\\ (Estimated) \end{tabular}\\
\midrule
Age & -- & 52.8 & 52.8 & 51.7 & 52.8\\
Sex (\%) & Male & 25.5 & 38.6 & 79.5 & 47.7\\
 & Female & 74.5 & 61.4 & 20.5 & 52.3\\
\addlinespace
Education (\%)& $<$HS & 3.6 & 6.6 & 12.2 & 9.1\\
 & HS & 19.9 & 34.8 & 46.8 & 34.7\\
 & College & 53.4 & 54.2 & 41 & 47\\
 & Graduate & 23 & 4.4 & 0 & 9.1\\
\addlinespace
Household income (\%)& 0-10k & 26.9 & 24.2 & 13.5 & 19.3\\
 & 10-20k & 25.4 & 17.3 & 12.8 & 18.4\\
 & $>$20k & 47.7 & 58.5 & 73.7 & 62.3\\
\addlinespace
Ethnicity (\%)& Puerto Rican & 96.4 & 96.3 & 94.9 & 94.9\\
 & Other & 3.6 & 3.7 & 5.1 & 5.1\\
\midrule
Urbanicity (\%)& Rural/In-between & 37.9 & 39.6 & 35.9 & 37.7\\
 & Urban & 62.1 & 60.4 & 64.1 & 62.3\\
\addlinespace
Smoking status (\%)& Never & 70.2 & 67.8 & 64.5 & 64.9\\
& Former & 18.7 & 19.2 & 20.6 & 22.4\\
 & Current & 11 & 13 & 14.8 & 12.6\\
\addlinespace
Drinking status (\%)& Never & 29.8 & 28.4 & 20 & 25.5\\
 & Former & 18.2 & 20.6 & 23.2 & 21.1\\
 & Current & 52 & 50.9 & 56.8 & 53.4\\
\addlinespace
Physical activity (\%)& Sedentary & 70.1 & 67.3 & 61.6 & 64.5\\
 & Active & 29.9 & 32.7 & 38.4 & 35.5\\
\addlinespace
Food security (\%)& Insecure & 18.1 & 18.1 & 15.4 & 15.9\\
& Secure & 81.9 & 81.9 & 84.6 & 84.1\\
\addlinespace
WIC SNAP use (\%)& No & 47.1 & 48.3 & 54.7 & 50.7\\
 & Yes & 52.9 & 51.7 & 45.3 & 49.3\\
\addlinespace
Depression (\%)& None/Mild & 70.2 & 69.8 & 76.3 & 71.4\\
 & Moderate/Severe & 29.8 & 30.2 & 23.7 & 28.6\\
\addlinespace
Anxiety (\%)& None/Mild & 79.6 & 77.8 & 82.3 & 77.8\\
 & Moderate/Severe & 20.4 & 22.2 & 17.7 & 22.2\\
Social support & -- & 27.4 & 27.6 & 27.9 & 27.9\\
Perceived stress & -- & 16.9 & 17 & 16.6 & 17\\
\bottomrule
\end{tabular}
\label{tab:weights_res}
\end{table}

\subsubsection{Dietary Behavior Patterns}

\begin{figure}[!htb]
    \centering
    \includegraphics[width=0.8\linewidth]{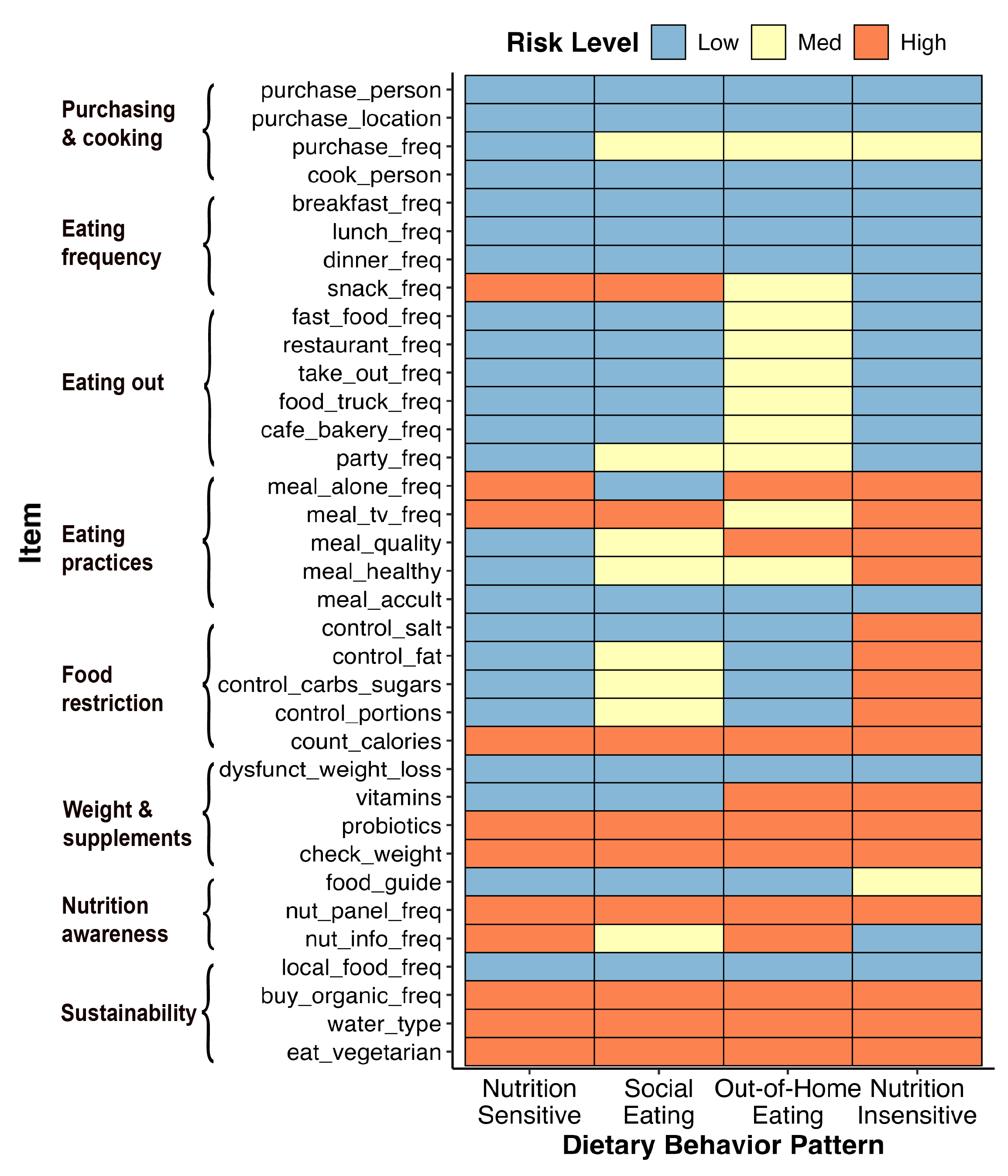}
    \caption{\textmd{Four dietary behavior patterns identified by the WOLCAN model among Puerto Rican adults aged 30 to 75. For each pattern, the highest probability (i.e., modal) risk level is displayed.}}
    \label{fig:WOLCAN_profiles}
\end{figure}

The WOLCAN procedure identified $\hat{K}=4$ dietary behavior patterns. The unweighted model (Appendix Figure 4) also identified 4 patterns, but included differences in most probable (i.e., modal) risk level for all patterns and, notably, failed to differentiate dieting practices across patterns. This illustrates the importance of incorporating weighting when eliciting patterns. In particular, we see the phenomenon with pattern mis-estimation that appeared in the simulations reflected here. For the unweighted model, the dieting practices variables have modal level set to low risk (i.e., frequent dieting practices) across all dietary behavior patterns. Since women are more likely to be in PROSPECT and are more likely to practice frequent dieting practices, regardless of the dietary behavior pattern, failure to include weights results in dietary behavior patterns that inflate the prevalence of frequent dieting practices due to the overrepresentation of women in the sample. 

Focusing on the patterns identified by WOLCAN, Figure \ref{fig:WOLCAN_profiles} displays each pattern's modal risk level for all dietary behavior variables. A more detailed breakdown of the risk level probabilities is provided in Appendix Figure 1. The four patterns can be characterized as: 1) nutrition-sensitive, 2) social eating, 3) out-of-home eating, and 4) nutrition-insensitive, with interpretation for the behavior variable risk levels aided by referencing the categorization scheme in Table \ref{tab:dietary_behavior}. The nutrition-sensitive pattern (DBP 1) consisted of individuals who tend to purchase groceries weekly, report high quality meals,  frequently modify meals to be healthier, and restrict intake of salt, fat, carbs/sugars, and portions. This pattern also involved a high frequency of snacking and infrequent seeking out of nutritional information from media sources. The social eating pattern (DBP 2) was unique in including a large proportion of individuals who rarely eat meals alone and also attend parties moderately frequently. Across other variables, this pattern was similar to the nutrition-sensitive pattern but included bi-weekly grocery purchasing, moderate meal quality, moderate (instead of high) frequency of dieting practices, and moderate nutrition-seeking from media sources. The out-of-home eating pattern (DBP 3) consisted of individuals who eat out the most often compared to all other patterns. They were likely to consume food from fast food establishments, restaurants, take out, cafes and bakeries, food trucks, and parties around 2-4 times per week. Snacking occurred moderately frequently and meal consumption in front of the TV occurred less frequently compared to other patterns. Out-of-home eaters also tended to report poor meal quality, but still modified meals to be healthier 2-4 times per week, as well as engaged in frequent dieting practices, similar to the nutrition-sensitive pattern. Lastly, the nutrition-insensitive pattern (DBP 4) consisted of individuals who reported poorer meal quality, infrequent health-conscious meals, none or minimal dieting practices, and lack of familiarity with the My Plate food guide. Compared to other patterns, nutrition-insensitive individuals also tended to snack less and more frequently seek out nutritional information from media sources such as TV, magazines, internet, and social media. 

Across all patterns, dietary behaviors that were uniformly observed at high modal risk level and could be improved through greater adoption included: counting calories, taking probiotics, checking weight, reading nutrition panels, buying organic, drinking tap water, and eating vegetarian. Behaviors that were low-risk across all patterns included: purchasing food directly, cooking food directly, purchasing food from supermarket chains, eating all three meals consistently throughout the week, eating mostly Puerto Rican meals rather than American meals, not participating in dysfunctional weight loss behaviors (e.g., vomiting), and frequently purchasing locally produced foods in Puerto Rico. 

Table \ref{tab:covs_across_dbh_res} compares the distribution of sociodemographic and behavioral variables across the four patterns. The nutrition-sensitive pattern was the most prevalent in both the PROSPECT sample and the population. It also had the oldest average age and was most represented among those with graduate level educational attainment, former smokers, and former drinkers. The social eating pattern had the highest proportion of female individuals, sedentary physical activity, never smokers, never drinkers, those who are food secure, and also those who receive WIC/SNAP food assistance. These individuals also had the most social support, which aligns with the shared dining behavior. The out-of-home eating pattern had the youngest average age and a high proportion of male individuals, those with a college or graduate-level educational attainment, active individuals, current smokers, and current drinkers. Notably, the out-of-home pattern had a much larger proportion of individuals who are not of Puerto Rican ethnicity, compared to the other patterns. They also tended to be high-income earners and non-users of WIC/SNAP food assistance, which aligns with the financial feasibility to eat out frequently. The nutrition-insensitive pattern had the highest proportion of individuals who are male, have lower levels of educational attainment, and have lower income, as well as a high proportion of former or current smokers or drinkers. Nutrition-insensitive individuals were also the most food insecure and had the highest levels of depressive symptoms, anxiety, lack of social support, and perceived stress, indicating an at-risk and vulnerable group. 

\begin{table}
    \caption{\textmd{Size and socio-behavioral distribution of derived dietary behavior patterns for adults aged 30 to 75 living in Puerto Rico. Estimates for $N$ use posterior samples of parameter $\mathbf{\pi}$. Column-wise mean and percentage estimates of socio-behavioral variables are calculated using estimated pseudo-weights.}}
    \scriptsize
    \centering
    \begin{tabular}{@{}llllllllll@{}}
    \toprule
    Variable & Level & DBP1 & DBP2 & DBP3 & DBP4 & Overall\\
    \midrule
    N: \% (posterior SD \%) & -- & 47.07 (22.60) & 23.90 (18.66) & 13.72 (7.28) & 15.32 (9.09) & \\
    n: \% & -- & 47.29 & 23.87 & 15.23 & 13.61 & \\
    \addlinespace
    Age: Mean & -- & 55.01 & 54.04 & 47.13 & 51.42 & 52.82\\
    Female: Mean &  & 0.57 & 0.66 & 0.40 & 0.35 & 0.52\\
    \addlinespace
    Education: \% & $<$HS & 10.10 & 6.02 & 6.99 & 13.11 & 9.14\\
     & HS & 31.56 & 41.71 & 26.97 & 42.10 & 34.75\\
     & College & 47.10 & 45.83 & 56.25 & 38.73 & 47.04\\
     & Graduate & 11.25 & 6.43 & 9.79 & 6.06 & 9.07\\
    \addlinespace
    Household income: \% & 0-10k & 19.49 & 21.59 & 13.43 & 21.99 & 19.30\\
     & 10-20k & 16.50 & 23.02 & 11.41 & 24.95 & 18.44\\
     & $>$20k & 64.01 & 55.39 & 75.16 & 53.06 & 62.26\\
     \addlinespace
    Ethnicity not PR: Mean & -- & 0.04 & 0.05 & 0.13 & 0.01 & 0.05\\
    Urban: Mean & -- & 0.60 & 0.64 & 0.64 & 0.65 & 0.62\\
    \addlinespace
    Physical activity: \% & Sedentary & 67.11 & 67.49 & 58.53 & 60.11 & 64.51\\
 & Active & 32.89 & 32.51 & 41.47 & 39.89 & 35.49\\
 \addlinespace
    Smoking status: \% & Never & 63.66 & 71.94 & 66.36 & 57.46 & 64.95\\
     & Former & 27.08 & 16.77 & 14.59 & 25.85 & 22.40\\
     & Current & 9.27 & 11.29 & 19.05 & 16.69 & 12.65\\
     \addlinespace
    Drinking status: \% & Never & 22.78 & 37.79 & 22.14 & 19.49 & 25.47\\
     & Former & 27.33 & 17.46 & 12.89 & 18.05 & 21.09\\
     & Current & 49.89 & 44.75 & 64.97 & 62.46 & 53.45\\
     \addlinespace
    Food secure: Mean & -- & 0.84 & 0.87 & 0.82 & 0.81 & 0.84\\
    Receive WIC/SNAP: Mean & -- & 0.51 & 0.56 & 0.35 & 0.51 & 0.49\\
    \addlinespace
    Depressive symptoms: Mean &  & 0.25 & 0.33 & 0.23 & 0.39 & 0.29\\
    Anxiety: Mean &  & 0.22 & 0.23 & 0.19 & 0.25 & 0.22\\
    \addlinespace
    Social support: Mean & -- & 27.17 & 29.22 & 29.04 & 26.68 & 27.88\\
    Perceived stress: Mean & -- & 17.21 & 15.43 & 17.35 & 17.97 & 16.97\\
    \bottomrule
    \end{tabular}
    \label{tab:covs_across_dbh_res}
\end{table}

\subsubsection{Association with Hypertension, Type 2 Diabetes, and Hypercholesterolemia}

\begin{figure}[!htb]
    \centering
    \includegraphics[width=0.8\linewidth]{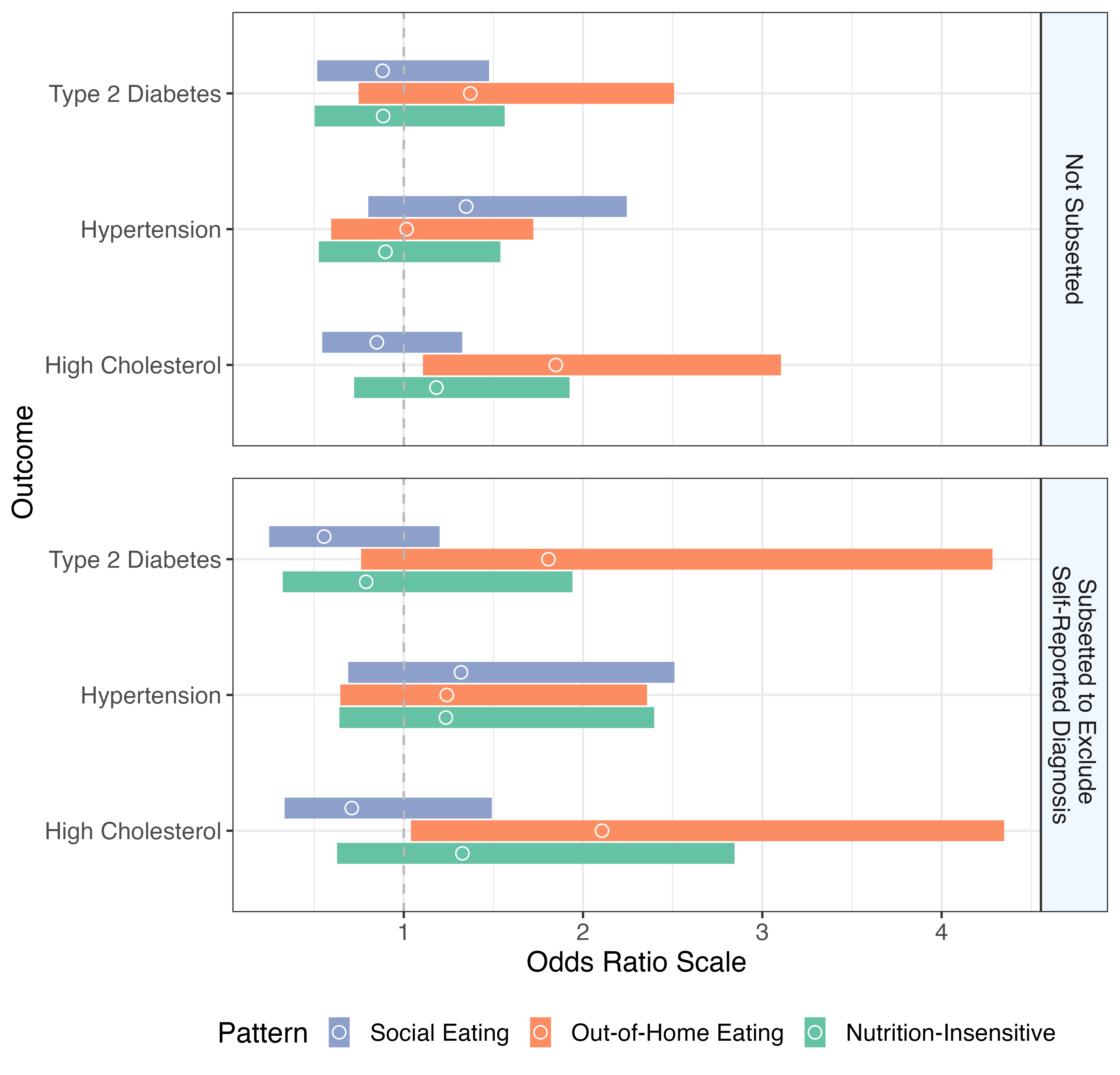}
    \caption{\textmd{Weighted logistic regression odds ratios and 95\% credible intervals (CI) for the outcomes of type 2 diabetes, hypertension, and hypercholesterolemia, comparing the social eating, out-of-home eating, and nutrition-insensitive dietary behavior patterns to the reference nutrition-sensitive pattern. Models were run in the full sample including individuals with self-reported or laboratory-based diagnoses (upper), as well as subsetted to exclude those with self-reported diagnoses to control for reverse causation (lower). All models adjusted for age, sex, educational attainment, annual household income, ethnicity, urbanicity, physical activity, smoking status, drinking status, food security, use of WIC/SNAP food assistance, social support, perceived stress, depression, anxiety, and the other two comorbidities (e.g., hypertension and hypercholesterolemia for type 2 diabetes). The dashed vertical line occurs at an odds ratio of 1.}}
    \label{fig:full_outcome_plot}
\end{figure}

Table \ref{tab:full_outcome} and Figure \ref{fig:full_outcome_plot} display the results of the weighted logistic regression models, which include the dietary behavior pattern assignments as covariates of interest and account for uncertainty in the pseudo-weights. For each outcome (type 2 diabetes, hypertension, and hypercholesterolemia), posterior regression estimates and 95\% credible intervals are provided for the odds ratios comparing individuals following the social eating, out-of-home eating, or nutrition-insensitive patterns compared to the reference nutrition-sensitive pattern, controlling for age, sex, educational attainment, annual household income, ethnicity, urbanicity, physical activity, smoking status, drinking status, food security, receipt of WIC/SNAP food assistance, social support, perceived stress, depressive symptoms, anxiety, and the other two outcomes of interest (e.g., hypertension and hypercholesterolemia for type 2 diabetes). 

Among individuals without a self-reported diagnosis (displayed in the lower panel), the out-of-home pattern had the highest odds of type 2 diabetes and hypercholesterolemia, while the social pattern had the lowest odds of type 2 diabetes and hypercholesterolemia. The out-of-home pattern was associated with an odds ratio (95\% credible interval) of 1.81 (0.76, 4.28) for type 2 diabetes and 2.11 (1.04, 4.35) for hypercholesterolemia, compared to the nutrition-sensitive pattern. All patterns were associated with a small increase in odds of hypertension as compared to the nutrition-sensitive pattern.  
 
Sensitivity analysis of the dietary patterns on the cardiometabolic outcomes indicated potential areas of reverse causality through dietary behavior change resulting from known diagnoses. Excluding individuals aware of their diagnosis led to higher odds of hypertension and hypercholesterolemia for the out-of-home and nutrition-insensitive patterns, compared to the nutrition-sensitive pattern, relative to the full sample that included both self-reported and laboratory-based diagnoses. A similar increase was observed for the odds of type 2 diabetes for the out-of-home pattern. Associations for the social eating showed the converse trend. Across all three outcomes, the odds ratios became more strongly protective after removing individuals aware of their diagnosis. 

\begin{table}[!htb]
    \caption{\textmd{Weighted logistic regression ORs and 95\% CIs for the outcomes of type 2 diabetes, hypertension, and hypercholesterolemia, comparing the social eating, out-of-home eating, and nutrition-insensitive dietary behavior patterns to the reference nutrition-sensitive pattern. Models were run in the full sample as well as subsetted to exclude those with self-reported diagnoses. The estimate for the reference level is the odds of the outcome when setting all variables to their reference or mean levels. Reference group: female, graduate education, household income $>20k$, Puerto Rican ethnicity, active, never smoker, never drinker, food secure, no WIC/SNAP use, none/mild depression, none/mild anxiety, no comorbidities. Abbreviations: OR = odds ratio, PP = posterior probability. Posterior probability is calculated as the probability of the parameter being greater than 1 or less than 1, depending on the direction of the effect.}}
    \label{tab:full_outcome}
    \scriptsize
    \centering
    \begin{tabular}{ll@{\hskip 0.2cm}ll@{\hskip 0.2cm}ll@{\hskip 0.2cm}l}
    \toprule
    & \multicolumn{2}{c}{Type 2 Diabetes} & \multicolumn{2}{c}{Hypertension} & \multicolumn{2}{c}{Hypercholesterolemia} \\
    Covariate & OR (95\% CI) & PP & OR (95\% CI) & PP & OR (95\% CI) & PP\\
    \midrule
    \multicolumn{7}{l}{Full Sample Model}\\
    \midrule
    Nutrition-Sensitive (Ref) & 0.02 (0.01, 0.05) & -- & 1.01 (0.53, 1.90) & -- & 0.37 (0.20, 0.68) & --\\
    Social Eating & 0.88 (0.52, 1.48) & 0.68 & 1.35 (0.80, 2.24) & 0.88 & 0.85 (0.54, 1.33) & 0.76\\
    Out-of-Home Eating & 1.37 (0.75, 2.51) & 0.85 & 1.02 (0.60, 1.72) & 0.52 & 1.85 (1.11, 3.10) & 0.99\\
    Nutrition-Insensitive & 0.88 (0.50, 1.56) & 0.66 & 0.90 (0.53, 1.54) & 0.65 & 1.18 (0.72, 1.93) & 0.75\\
    \midrule
    \multicolumn{7}{l}{Subsetted to Exclude Self-Reported Diagnosis}\\
    \midrule
    Nutrition-Sensitive (Ref) & 0.00 (0.00,  0.01) & -- & 0.27 (0.12, 0.61) & -- & 0.09 (0.03,  0.26) & --\\
    Social Eating & 0.56 (0.25,  1.20) & 0.93 & 1.32 (0.69, 2.51) & 0.80 & 0.71 (0.34,  1.49) & 0.82\\
    Out-of-Home Eating & 1.81 (0.76,  4.28) & 0.91 & 1.24 (0.65, 2.36) & 0.75 & 2.11 (1.04,  4.35) & 0.98\\
    Nutrition-Insensitive & 0.79 (0.32,  1.94) & 0.70 & 1.24 (0.64, 2.40) & 0.73 & 1.33 (0.63,  2.85) & 0.78\\
    \bottomrule
    \end{tabular}
\end{table}

\section{Discussion}\label{sec:discussion}
The proposed WOLCAN (weighted overfitted latent class analysis for non-probability samples) is a two-step method. In the first step, it estimates the selection mechanism responsible for generating a non-probability sample (NPS) using Bayesian additive regression trees (BART). In the second step, it performs a weighted model-based clustering technique to uncover latent patterns in a data-driven manner, ensuring generalizability to a broader population. Uncertainty from the first step is propagated under a Bayesian framework that enables classification in the original sample to also be recovered. Simulation studies highlighted the advantages of using BART to incorporate selection information from a reference probability sample (PS), enabling the generation of pseudo-weights for the NPS. Additionally, the WOLCAN model effectively recovered the true population parameters with minimal bias while obtaining nominal coverage in posterior intervals across various settings, outperforming the unweighted latent class model.

In the data application to the PROSPECT NPS, pseudo-weights were generated using the PRCS as a reference, which helped identify key underrepresented subgroups in the PROSPECT sample. These included individuals who were male, had lower educational attainment, had higher income, were food secure, did not receive WIC/PAN, and/or exhibited none/mild depressive symptoms. Using these pseudo-weights, WOLCAN identified four distinct dietary behavior patterns among adults in Puerto Rico aged 30 to 75 years: 1) nutrition-sensitive, 2) social eating, 3) out-of-home eating, and 4) nutrition-insensitive. Among these groups, the nutrition-insensitive group was more prevalent among individuals experiencing food insecurity, low education, and poor psychosocial health, denoting a vulnerable group. However, despite the presence of this at-risk pattern, the strongest adverse associations with the three cardiometabolic outcomes of interest -- type 2 diabetes, hypertension, and hypercholesterolemia -- were observed in the out-of-home eating pattern. Notably, this group exhibited a significantly higher risk of hypercholesterolemia even after adjusting for sociodemographic and behavioral confounder variables. Individuals following an out-of-home eating pattern were more likely to be male, younger, of non-Puerto Rican ethnicity, with higher educational attainment, and/or with higher income. For hypertension, the nutrition-sensitive pattern was associated with lower odds of developing the condition, suggesting that enhancing nutritional knowledge may be a key preventive strategy. The social eating pattern, characterized by frequent shared meals and moderate dieting practices, showed a protective association for type 2 diabetes and hypercholesterolemia, even when compared to the nutrition-sensitive group. Demographically, individuals in the social eating group were more likely to be older females, indicating potential behavioral and social advantages in dietary habits. 

The findings presented of this analysis align with previous research on dietary behaviors in Puerto Rico, though most prior studies have primarily focused on the San Juan area. \citet{lopez2023self}  showed that nutrition-sensitive behaviors (e.g., preparing healthier meals and controlling salt, fat, sugar, and portions) are linked to better diet quality and lower cardiometabolic risk. Similarly, \citet{bezares2023consumption} identified out-of-home eating, particularly of commercially prepared foods, to be associated with poorer diet quality, including higher sodium, added sugars, and saturated fat intake. Consistent with our findings, out-of-home eating was more prevalent among younger and higher income individuals \citep{colon2013socio}. However, our results also indicate that this pattern is more common among males and high-income groups, contrasting with previous research that associated it with females and low-income populations. Additionally, evidence suggests that positive social networks and strong family systems contribute to higher diet quality among Latinos living in the United States \citep{schmied2014family, miller2023food}, a trend that is reflected in our findings, particularly among older females. 

This work has demonstrated the value of the WOLCAN model in identifying and characterizing latent patterns in studies that rely on non-probability samples. However, several limitations should be acknowledged. First, the model employs complete case analysis without incorporating strategies to handle missing data. Second, the weighted outcome regression was applied as a separate step following the derivation of the dietary behavior patterns. This two-step approach may lead to attenuated effects due to unaccounted measurement error in the pattern derivation stage \citep{muthen1999finite, elliott2020methods, bray2015eliminating}, which likely contributed to the wide 95\% credible intervals observed in the results. Third, while WOLCAN adjusts for unequal probability of inclusion, it does not yet account for clustering effects. Fourth, a crude weight trimming method is used to remove extreme weights from greatly inflating the variance. Alternative approaches that threshold units based on optimal variance considerations may be more efficient \citep{savitsky2024thresholding}. Lastly, generalizability of the results depends on the selection, availability, and quality of the variables used to model selection into the non-probability sample. Future research can address these limitations by integrating missing data handling techniques, refining the modeling approach to reduce measurement error, incorporating clustering adjustments, more efficiently handling extreme weights, and enhancing selection modeling strategies to improve applicability and accuracy.

\section*{Acknowledgments}
The authors sincerely thank Elizabeth Petit, Jonathan Orozco, Karla Medina, Carlos Rios-Bedoy, and Sigrid Mendoza for their invaluable support with the PROSPECT study data. The authors are also grateful to Trivellore Raghunathan and Yajuan Si for their thoughtful feedback on earlier versions of this work. 

\section*{Funding}
This research was supported by the Harvard Data Science Initiative Bias-Squared Program and the 
National Heart Lung and Blood Institute (NHLBI K01HL166442-01).

\section{Competing interests}
No competing interests are declared.

\section*{Data availability}
Code for replicating the simulations and data analyses in this paper is available on GitHub at \url{https://github.com/smwu/WOLCAN}.


\bibliographystyle{abbrvnat}
\bibliography{reference}

\clearpage 
\section*{Supplementary Materials}
\setcounter{table}{0}
\setcounter{figure}{0}

\subsection*{Appendix A: Simulation Study}
\subsubsection*{Simulation parameter values for multinomial logistic regression generating $X_{ij}$}\label{app:betas}

For the latent class analysis simulation design, the manifest categorical variables $\mathbf{x}_i = (x_{i1},\ldots, x_{iJ})$, with $J=30$ and $R=4$ category levels, are generated to depend on $c_i$, $a_{i1}$, and $a_{i3}$ through $J$ multinomial logistic regression formulations. Since the latent class assignments take on values $c_i\in \{1,2,3\}$, the  multinomial logistic regression formulation for each item $j$ is given by:
\begin{align}
    x_{ij}|c_i, a_{i1}, a_{i3} &\sim \text{Multinomial}(1, \theta_{ij1},\ldots, \theta_{ijR})\\
    \text{where } \eta_{ijr} &:= \log\bigg(\frac{\theta_{ijr}}{\theta_{ij1}}\bigg) = \beta_{jr0} + \beta_{jr1}I(c_i=2) + \beta_{jr2}I(c_i=3) + \beta_{jr3}a_{i1} + \beta_{jr4}a_{i3} \nonumber \\
    &\qquad + \beta_{jr5}I(c_i=2)a_{i1} + \beta_{jr6}I(c_i=3)a_{i1} \text{ for } r=1,\ldots, R, \\
    \text{with } \theta_{ij1} &= \frac{1}{1+\sum\limits_{r=2}^R e^{\eta_{ijr}}}\text{ and }\
    \theta_{ijr} = \frac{e^{\eta_{ijr}}}{1+\sum\limits_{r=2}^R e^{\eta_{ijr}}} \text{ for } r= 2,\ldots, R. 
\end{align}

For $j=1$ to $j=2$, we have
$$
\begin{pmatrix}\beta_{j10} &\beta_{j11} & \beta_{j12} & \beta_{j13} & \beta_{j14} & \beta_{j15}  & \beta_{j16}\\ 
\beta_{j20} &\beta_{j21} & \beta_{j22} & \beta_{j23} & \beta_{j24} & \beta_{j25} & \beta_{j26}\\ \beta_{j30} &\beta_{j31} & \beta_{j32} & \beta_{j33} & \beta_{j34} & \beta_{j35} & \beta_{j36} \\ \beta_{j40} &\beta_{j41} & \beta_{j42} & \beta_{j43} & \beta_{j44} & \beta_{j45} & \beta_{j46}\end{pmatrix} = 
\begin{pmatrix}0 &0 & 0 & 0 &0 &0 &0\\ -2.833 & 2.833 & 2.833 & 0 & 0.5 & 0 & 0\\ -2.833 & 2.833 & 5.666 & 0 & 0 & 0 & 0 \\ -2.833 & 5.666 & 2.833 & 0 & 0 & 0 & 0\end{pmatrix}. 
$$
For $j=3$ to $j=6$, we have
$$
\begin{pmatrix}\beta_{j10} &\beta_{j11} & \beta_{j12} & \beta_{j13} & \beta_{j14} & \beta_{j15}  & \beta_{j16}\\ 
\beta_{j20} &\beta_{j21} & \beta_{j22} & \beta_{j23} & \beta_{j24} & \beta_{j25} & \beta_{j26}\\ \beta_{j30} &\beta_{j31} & \beta_{j32} & \beta_{j33} & \beta_{j34} & \beta_{j35} & \beta_{j36} \\ \beta_{j40} &\beta_{j41} & \beta_{j42} & \beta_{j43} & \beta_{j44} & \beta_{j45} & \beta_{j46}\end{pmatrix} = 
\begin{pmatrix}0 &0 & 0 & 0 &0 &0 &0\\ -2.833 & 2.833 & 2.833 & 0 & 0 & 0 & 0\\ -2.833 & 2.833 & 5.666 & 0 & 0 & 0 & 0 \\ -2.833 & 5.666 & 2.833 & 0 & 0 & 0 & 0\end{pmatrix}. 
$$
For $j=7$ to $j=9$, we have
$$
\begin{pmatrix}\beta_{j10} &\beta_{j11} & \beta_{j12} & \beta_{j13} & \beta_{j14} & \beta_{j15}  & \beta_{j16}\\ 
\beta_{j20} &\beta_{j21} & \beta_{j22} & \beta_{j23} & \beta_{j24} & \beta_{j25} & \beta_{j26}\\ \beta_{j30} &\beta_{j31} & \beta_{j32} & \beta_{j33} & \beta_{j34} & \beta_{j35} & \beta_{j36} \\ \beta_{j40} &\beta_{j41} & \beta_{j42} & \beta_{j43} & \beta_{j44} & \beta_{j45} & \beta_{j46}\end{pmatrix} = 
\begin{pmatrix}0 &0 & 0 & 0 &0 &0 &0\\ -2.833 & 5.666 & 2.833 & 0 & 0 & 0 & 0\\ -2.833 & 2.833 & 5.666 & 0 & 0 & 0 & 0 \\ -2.833 & 2.833 & 2.833 & 0 & 0 & 0 & 0\end{pmatrix}. 
$$
For $j=10$ to $j=15$, we have
$$
\begin{pmatrix}\beta_{j10} &\beta_{j11} & \beta_{j12} & \beta_{j13} & \beta_{j14} & \beta_{j15}  & \beta_{j16}\\ 
\beta_{j20} &\beta_{j21} & \beta_{j22} & \beta_{j23} & \beta_{j24} & \beta_{j25} & \beta_{j26}\\ \beta_{j30} &\beta_{j31} & \beta_{j32} & \beta_{j33} & \beta_{j34} & \beta_{j35} & \beta_{j36} \\ \beta_{j40} &\beta_{j41} & \beta_{j42} & \beta_{j43} & \beta_{j44} & \beta_{j45} & \beta_{j46}\end{pmatrix} = 
\begin{pmatrix}0 &0 & 0 & 0 &0 &0 &0\\ -2.833 & 5.666 & 2.833 & 0 & 0 & 0 & 0\\ -2.833 & 2.833 & 2.833 & 0 & 0 & 0 & 0 \\ -2.833 & 2.833 & 5.666 & 0 & 0 & 0 & 0\end{pmatrix}. 
$$
For $j=16$ to $j=21$, we have
$$
\begin{pmatrix}\beta_{j10} &\beta_{j11} & \beta_{j12} & \beta_{j13} & \beta_{j14} & \beta_{j15}  & \beta_{j16}\\ 
\beta_{j20} &\beta_{j21} & \beta_{j22} & \beta_{j23} & \beta_{j24} & \beta_{j25} & \beta_{j26}\\ \beta_{j30} &\beta_{j31} & \beta_{j32} & \beta_{j33} & \beta_{j34} & \beta_{j35} & \beta_{j36} \\ \beta_{j40} &\beta_{j41} & \beta_{j42} & \beta_{j43} & \beta_{j44} & \beta_{j45} & \beta_{j46}\end{pmatrix} = 
\begin{pmatrix}0 &0 & 0 & 0 &0 &0 &0\\ 0 & 2.833 & 0 & 0 & 0 & 0 & 0\\ 2.833 & -2.833 & -2.833 & 0 & 0 & 0 & 0 \\ 0 & 0 & 2.833 & 0 & 0 & 0 & 0\end{pmatrix}. 
$$
For $j=22$ to $j=28$, we have
$$
\begin{pmatrix}\beta_{j10} &\beta_{j11} & \beta_{j12} & \beta_{j13} & \beta_{j14} & \beta_{j15}  & \beta_{j16}\\ 
\beta_{j20} &\beta_{j21} & \beta_{j22} & \beta_{j23} & \beta_{j24} & \beta_{j25} & \beta_{j26}\\ \beta_{j30} &\beta_{j31} & \beta_{j32} & \beta_{j33} & \beta_{j34} & \beta_{j35} & \beta_{j36} \\ \beta_{j40} &\beta_{j41} & \beta_{j42} & \beta_{j43} & \beta_{j44} & \beta_{j45} & \beta_{j46}\end{pmatrix} = 
\begin{pmatrix}0 &0 & 0 & 0 &0 &0 &0\\ 0 & 2.833 & -2.833 & 0 & 0 & 0 & 0\\ 2.833 & -2.833 & -5.666 & 0 & 0 & 0 & 0 \\ 0 & 0 & -2.833 & 0 & 0 & 0 & 0\end{pmatrix}. 
$$
For $j=29$ to $j=30$, we have
$$
\begin{pmatrix}\beta_{j10} &\beta_{j11} & \beta_{j12} & \beta_{j13} & \beta_{j14} & \beta_{j15}  & \beta_{j16}\\ 
\beta_{j20} &\beta_{j21} & \beta_{j22} & \beta_{j23} & \beta_{j24} & \beta_{j25} & \beta_{j26}\\ \beta_{j30} &\beta_{j31} & \beta_{j32} & \beta_{j33} & \beta_{j34} & \beta_{j35} & \beta_{j36} \\ \beta_{j40} &\beta_{j41} & \beta_{j42} & \beta_{j43} & \beta_{j44} & \beta_{j45} & \beta_{j46}\end{pmatrix} = 
\begin{pmatrix}0 &0 & 0 & 0 &0 &0 &0\\ 0 & 2.833 & -2.833 & 0 & 0 & 2 & -1\\ 2.833 & -2.833 & -5.666 & 2 & 0 & -2 & -2 \\ 0 & 0 & -2.833 & 0 & 0 & 0 & -1. 
\end{pmatrix}
$$

\clearpage
\subsection*{Appendix B: Data Application}
\subsubsection*{CONSORT diagram detailing inclusion and exclusion criteria for PROSPECT}\label{app:consort_PROSPEECT}
\vspace{1em}
\begin{singlespacing}
\begin{tikzpicture}%
  [data/.style=
    {draw,minimum height=1cm,minimum width=3cm,align=center, anchor=north west},
   filter/.style=
    {draw,minimum height=1cm,minimum width=3cm,align=center,fill=gray!30},
   database/.style=
    {draw,minimum height=1cm,minimum width=3cm,align=center},
   flow/.style={thick,-stealth},
   apply/.style={}
  ]
    \small
\node (start) [data] {\begin{tabular}{l} Total PROSPECT participants (n=1700)\end{tabular}};
\node[filter, right=of start] (excl) {Exclusions};
\node (xdata) [data, yshift=-2cm, below of=start, text width=4.5cm] {\begin{tabular}{l}Complete dietary \\ behavior data (n=1550)\end{tabular}};
\draw[flow] (start) -- coordinate(d1d2) (xdata);
\node[filter, text width=4cm] (f1) at (d1d2-|excl) {\begin{tabular}{l} Missing dietary behavior \\ data (n=150)\end{tabular}};
\draw[apply] (d1d2) -- (f1);
\node (xycovdata) [data, yshift=-2cm, below of=xdata, text width=4.5cm] {\begin{tabular}{l}Complete data for all \\variables (n=1353)\end{tabular}};
\draw[flow] (xdata) -- coordinate(d2d3) (xycovdata);
\node[filter, text width=4cm] (f2) at (d2d3-|excl) {\begin{tabular}{l} Missing covariate or \\ outcome data (n=197)\end{tabular}};
\draw[apply] (d2d3) -- (f2);
\node (subset) [data, yshift=-2cm, below of=xycovdata, xshift=3cm, text width=5cm] {\begin{tabular}{l} Subsetted analysis excluding \\ self-reported diagnosis: \\ 
\hspace{0.2em} Hypertension (n=1126)\\
\hspace{0.2em} Type 2 diabetes (n=737)\\
\hspace{0.2em} Hypercholesterolemia (n=838)\end{tabular}};
\node (fullsamp) [data, yshift=-1.32cm, below of=xycovdata, xshift=-3cm, text width=5cm] {\begin{tabular}{l}Full sample analysis (n=1353)\end{tabular}};
\draw[flow] (xycovdata) -- (subset);
\draw[flow] (xycovdata) -- (fullsamp);
\end{tikzpicture}
    
\end{singlespacing}

\subsubsection*{CONSORT Diagram Detailing Inclusion and Exclusion Criteria for PRCS}\label{app:consort_PRCS}
\vspace{1em}
\begin{singlespacing}
    
\begin{tikzpicture}%
  [data/.style=
    {draw,minimum height=1cm,minimum width=3cm,align=center, anchor=north west},
   filter/.style=
    {draw,minimum height=1cm,minimum width=3cm,align=center,fill=gray!30},
   database/.style=
    {draw,minimum height=1cm,minimum width=3cm,align=center},
   flow/.style={thick,-stealth},
   apply/.style={}
  ]
    \small
\node (start) [data] {\begin{tabular}{l} Total PRCS participants (n=132,642)\end{tabular}};
\node[filter, right=of start] (excl) {Exclusions};
\node (xdata) [data, yshift=-2cm, below of=start, text width=5cm] {\begin{tabular}{l}Complete  data (n=113,229)\end{tabular}};
\draw[flow] (start) -- coordinate(d1d2) (xdata);
\node[filter, text width=5cm] (f1) at (d1d2-|excl) {\begin{tabular}{l} Missing data (n=19,413)\end{tabular}};
\draw[apply] (d1d2) -- (f1);
\node (xdataage) [data, yshift=-2cm, below of=xdata, text width=5cm] {\begin{tabular}{l}Complete data for adults \\ aged 30-75 years (n=77,907)\end{tabular}};
\draw[flow] (xdata) -- coordinate(d2d3) (xdataage);
\node[filter, text width=5cm] (f2) at (d2d3-|excl) {\begin{tabular}{l} Age not within 30-75 \\ year range (n=35,322)\end{tabular}};
\draw[apply] (d2d3) -- (f2);
\node (fullsamp) [data, yshift=-2cm, below of=xdataage, text width=5.5cm] {\begin{tabular}{l} Sample size for pseudo-weight\\ estimation (n=77,907)\end{tabular}};
\draw[flow] (xdataage) -- (fullsamp);
\end{tikzpicture}

\end{singlespacing}

\clearpage
\subsubsection*{Results for the Proposed WOLCAN Model}

\noindent\textbf{Appendix Table~\ref{app:weights_res_validation}: Distribution of Selection Variables in the Sample and Population}

\begin{table}[H]
    \centering
    \centering
\small
\begin{tabular}{llcll}
\hline
\begin{tabular}[c]{@{}l@{}}Variable\\ \end{tabular} &
\begin{tabular}[c]{@{}l@{}}Level\\ \end{tabular} &
\begin{tabular}[c]{@{}c@{}}PROSPECT\\ Sample\end{tabular} &
\begin{tabular}[c]{@{}l@{}}Population\\ (WOLCAN)\end{tabular} &
\begin{tabular}[c]{@{}l@{}}Population \\ (PRCS)\end{tabular} \\ \hline
Age                    & --               & 52.8 & 52.8 & 52.4 \\
\addlinespace
Sex (\%)               & \textcolor{black}{Male}             & \textcolor{black}{25.5} & 47.7 & 46.5 \\
& Female           & 74.5 & 52.3 & 53.5 \\
\addlinespace
Education (\%)         & \textcolor{black}{$<$HS}            & \textcolor{black}{3.6}  & 9.1  & 10.6 \\
& \textcolor{black}{HS}               & \textcolor{black}{19.9} & 34.7 & 35.6 \\
& College          & 53.4 & 47   & 44.9 \\
& Graduate         & 23   & 9.1  & 8.9  \\
\addlinespace
Household income (\%)  & 0-10k            & 26.9 & 19.3 & 18.3 \\
& 10-20k           & 25.4 & 18.4 & 18.8 \\
& \textcolor{black}{$>$20k}           & \textcolor{black}{47.7} & 62.3 & 62.9 \\
\addlinespace
Ethnicity (\%)         & Puerto Rican     & 96.4 & 94.9 & 94.7 \\
& \textcolor{black}{Other}            & \textcolor{black}{3.6}  & 5.1  & 5.3  \\ \hline
\end{tabular}
    \caption{\textmd{Distribution of selection variables among all individuals in the PROSPECT sample, among all those in the target population using weights estimated with the WOLCAN model, and among all those in the population using the known survey weights from PRCS.  Means are reported for continuous variables, and column percentages are reported for categorical variables.}}
    \label{app:weights_res_validation}
\end{table}

\begin{figure}[H]
    \centering
    \includegraphics[width=\textwidth]{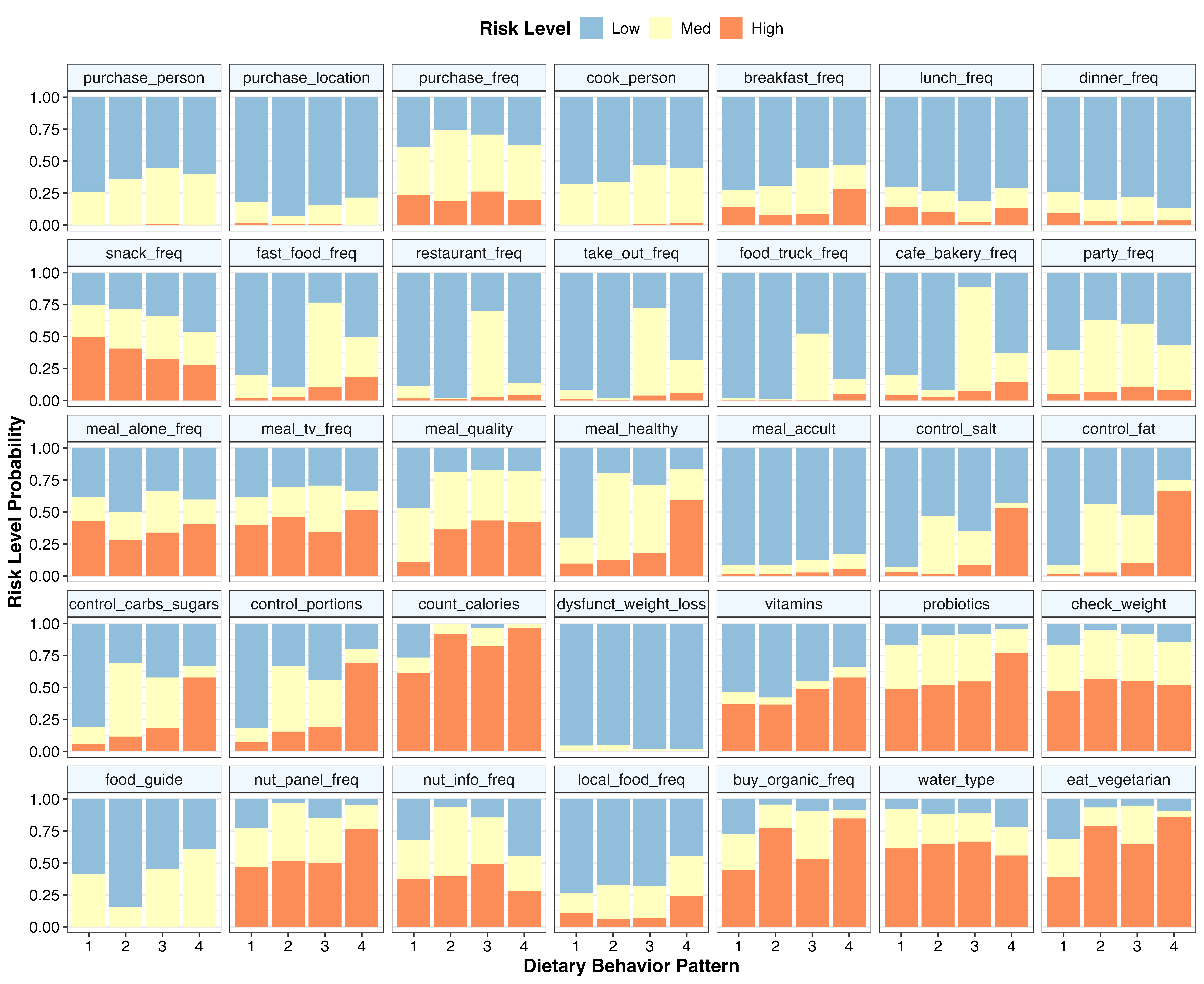}
    \caption{\textmd{Detailed breakdown of risk level probabilities by dietary behavior pattern for each dietary behavior variable. Risk levels are categorized as low, medium, and high.}}
    \label{app:wolcan_pattern_probs_plot}
\end{figure}

\clearpage
\noindent\textbf{Appendix Figure~\ref{app:wolcan_pattern_probs_plot}: WOLCAN Dietary Behavior Pattern Risk Level Probabilities}

\clearpage
\subsubsection*{Results for the Unweighted Bayesian LCA Model}
\begin{figure}[H]
    \centering
    \includegraphics[width=0.6\textwidth]{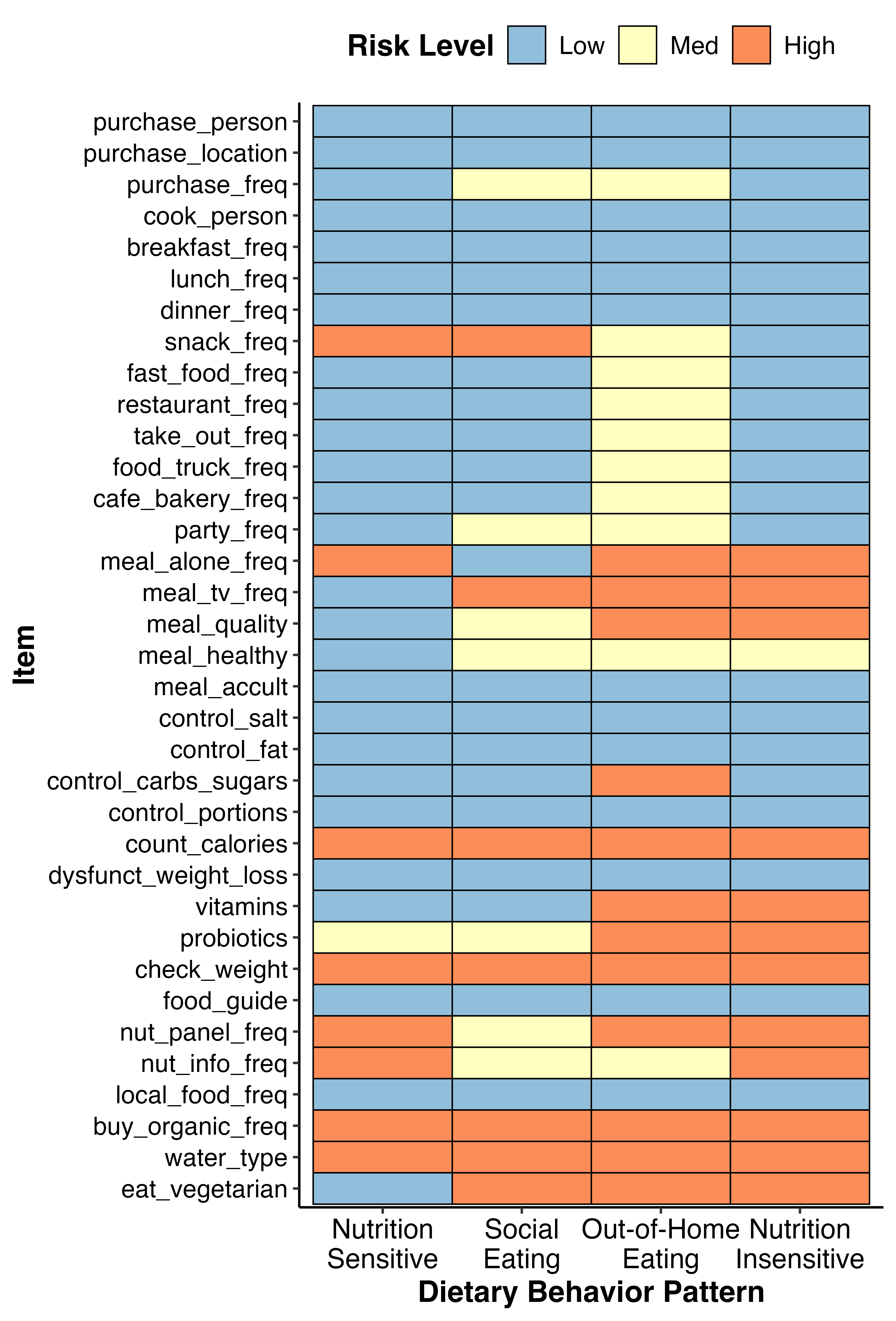}
    \caption{\textmd{Four dietary behaviors patterns identified by the unweighted Bayesian LCA model among adults aged 30 to 75 living in Puerto Rico. Risk levels are categorized as low, medium, and high. For each pattern, risk level of each dietary behavior variable is colored according to the modal risk level (i.e., $\text{argmax}_r \theta_{jkr}$ for $r=1,\ldots, 3$, $j = 1,\ldots, 35$, $k = 1,\ldots, 4$). Patterns are labeled based on the most similar analogous pattern from the WOLCAN results.}}
    \label{app:wolcan_unwt_pattern_profiles_plot}
\end{figure}

\noindent\textbf{Appendix Figure~\ref{app:wolcan_unwt_pattern_profiles_plot}: Unweighted Bayesian LCA Dietary Behavior Patterns}

\begin{figure}[H]
    \centering
    \includegraphics[width=\textwidth]{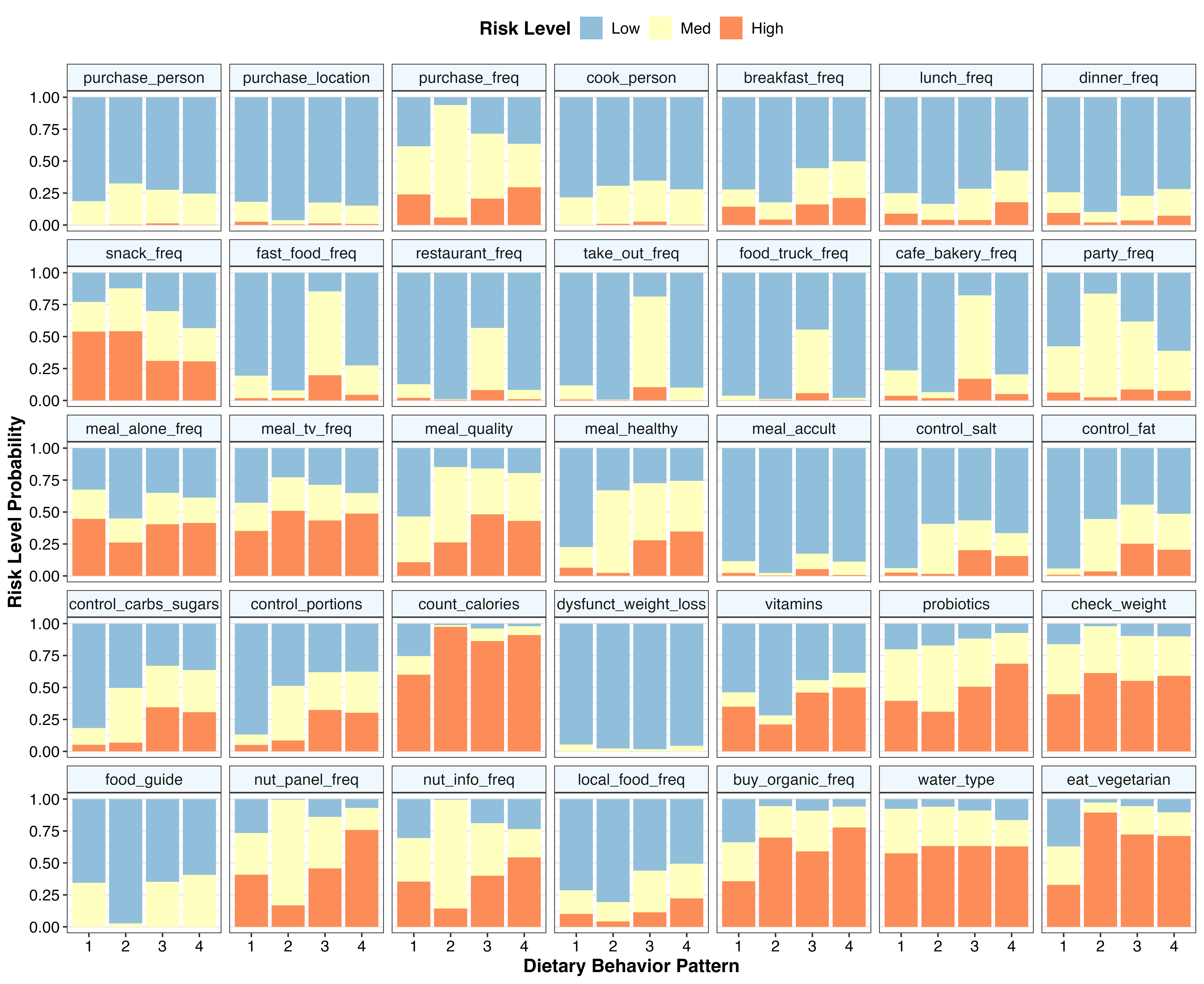}
    \caption{\textmd{Detailed breakdown of risk level probabilities by dietary behavior pattern for each dietary behavior variable. Risk levels are categorized as low, medium, and high.}}
    \label{app:wolcan_unwt_pattern_probs_plot}
\end{figure}

\clearpage
\noindent\textbf{Appendix Figure~\ref{app:wolcan_unwt_pattern_probs_plot}: Unweighted Bayesian LCA Dietary Behavior Pattern Risk Level Probabilities}

\begin{figure}[H]
    \centering
    \includegraphics[width=\linewidth]{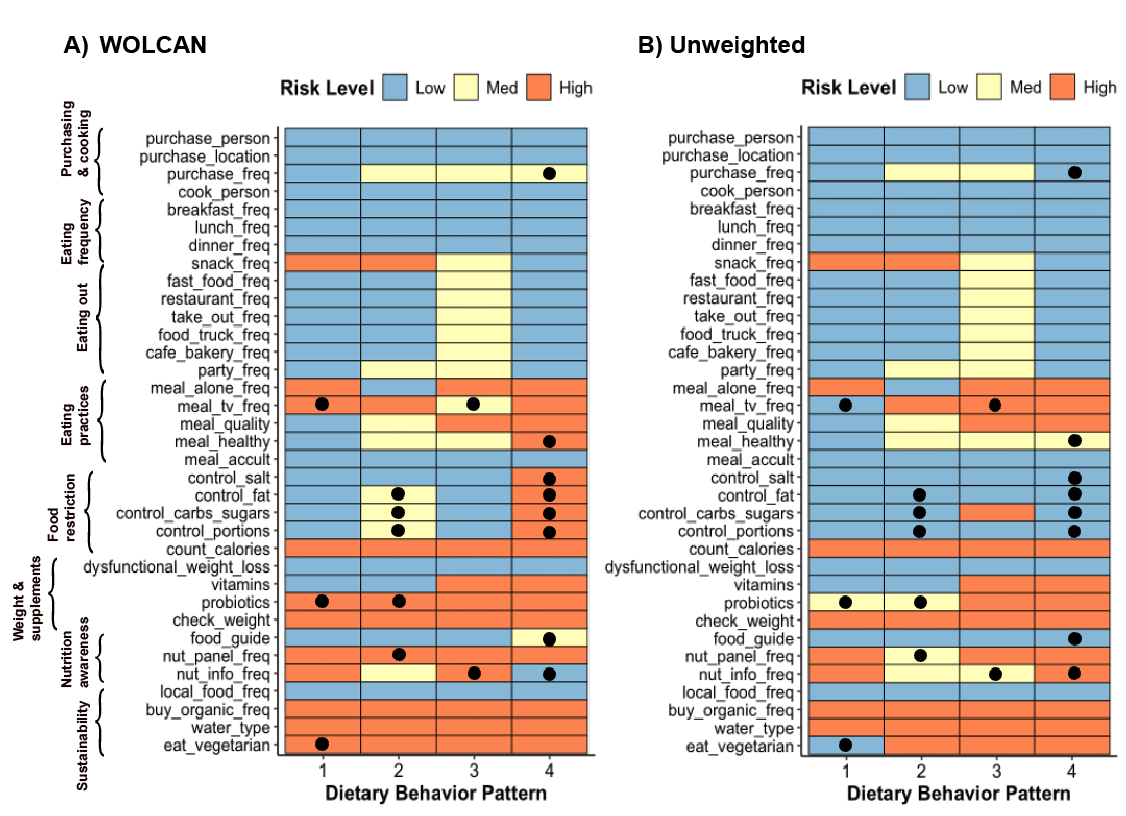}
    \caption{\textmd{Dietary behavior patterns elicited comparing the A) WOLCAN model and the B) unweighted Bayesian LCA model.}}
    \label{app:wolcan_profile_comparison}
\end{figure}

\clearpage
\noindent\textbf{Appendix Figure~\ref{app:wolcan_profile_comparison}: Comparison of WOLCAN and Unweighted Dietary Behavior Patterns}

\begin{figure}[!htb]
    \centering
    \includegraphics[width = 0.8\linewidth]{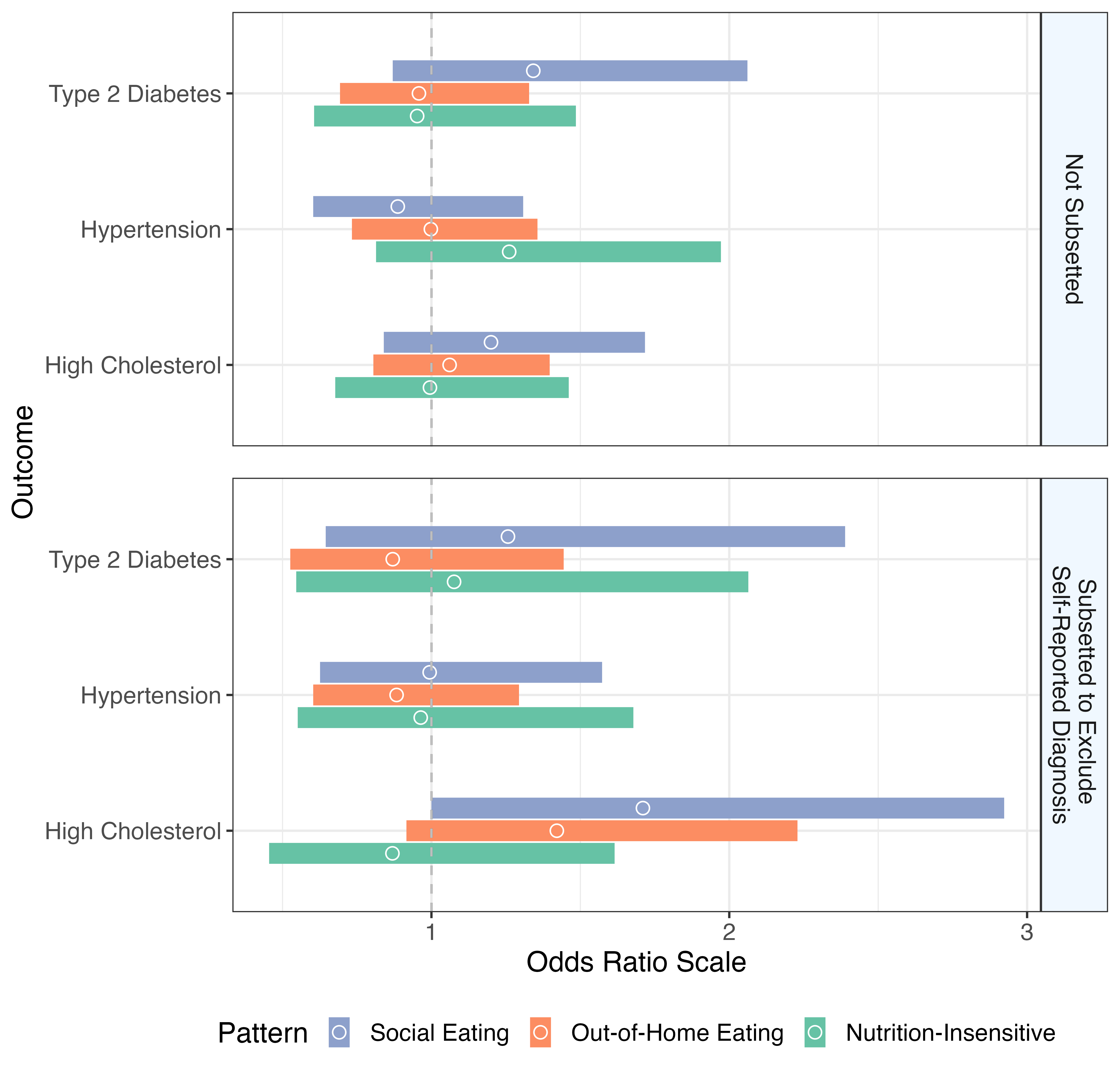}
    \caption{\textmd{Unweighted logistic regression odds ratios and 95\% confidence intervals (CI) for the outcomes of type 2 diabetes, hypertension, and hypercholesterolemia, comparing the \textit{unweighted} analogous social eating, out-of-home eating, and nutrition-insensitive dietary behavior patterns to the analogous reference nutrition-sensitive pattern. Models were run in the full sample including individuals with self-reported or laboratory-based diagnoses (upper), as well as subsetted to exclude those with self-reported diagnoses to control for reverse causation (lower). All models adjusted for age, sex, educational attainment, annual household income, ethnicity, urbanicity, physical activity, smoking status, drinking status, food security, use of WIC/SNAP food assistance, social support, perceived stress, depression, anxiety, and the other two comorbidities (e.g., hypertension and hypercholesterolemia for type 2 diabetes). The dashed vertical line occurs at an odds ratio of 1.}}
    \label{app:wolcan_unwt_full_outcome_plot}
\end{figure}

\clearpage
\noindent\textbf{Appendix Figure~\ref{app:wolcan_unwt_full_outcome_plot}: Unweighted Logistic Regression Results}

\end{document}